\documentclass[12pt,preprint]{aastex}
\shorttitle{$\alpha$-SBF}
\shortauthors{H.-c. Lee et al.}

\begin{document}

\title{Effects of Alpha-Element Enhancement and the Thermally 
Pulsing-Asymptotic Giant Branch 
on Surface Brightness Fluctuation Magnitudes and Broadband Colors}

\author{Hyun-chul Lee}
\affil{Department of Physics and Astronomy, Washington State University, 
Pullman, WA 99164-2814; and Department of Physics and Geology, 
The University of Texas - Pan American, Edinburg, TX 78539}
\email{leeh@utpa.edu}
\author{Guy Worthey}
\affil{Department of Physics and Astronomy, Washington State University, 
Pullman, WA 99164-2814}
\author{John P. Blakeslee}
\affil{NRC Herzberg Institute of Astrophysics, Victoria, BC, V9E 2E7, Canada}

\begin{abstract}

We investigate the effects of $\alpha$-element enhancement 
and the thermally pulsing-asymptotic giant branch (TP-AGB) stars 
on the surface brightness fluctuation (SBF) magnitudes and broadband 
colors of simple stellar populations and compare to the empirical 
calibrations.  We consider a broad range of ages and metallicities 
using the recently updated Teramo BaSTI isochrones.  
We find that the $\alpha$-element enhanced $I$-band SBF magnitudes 
are about 0.35 mag brighter and their integrated $V - I$ colors are 
about 0.02 mag redder, 
mostly because of oxygen enhancement effects on the upper red giant 
branch and asymptotic giant branch.  We also demonstrate, using both 
the Teramo BaSTI and Padova isochrones, the acute sensitivity of SBF 
magnitudes to the presence of TP-AGB stars, particularly in the near-IR, 
but in the $I$-band as well.  Empirical SBF trends therefore hold great 
promise for constraining this important but still highly uncertain 
stage of stellar evolution. In a similar vein, 
non-negligible disparities are found among several different models 
available in the literature due to intrinsic model uncertainties.  

\end{abstract}


\section{Introduction}

The surface brightness fluctuation (SBF) method, which measures 
the intrinsic pixel-to-pixel intensity variance in a galaxy image, 
is widely used as one of the most powerful distance indicators 
as well as a useful tool for probing stellar populations in the 
integrated light of early-type galaxies and spiral bulges.  
It is now a well-known fact that SBF magnitudes vary as a function 
of galaxy colors (e.g., Tonry et al. 2001; Jensen et al. 2003).  
For instance, empirical relations show that galaxies with redder 
$V - I$ colors have fainter $I$-band SBF magnitudes (e.g., Tonry 1991;
Tonry et al. 1997, 2001).  Moreover, there have been some suggestions 
that bluer dwarf elliptical galaxies have a shallower slope in the 
$V - I$  vs.\ $I$-band SBF magnitude diagram compared to the redder, 
massive galaxies (e.g., Blakeslee et al.\ 2001; Mei et al. 2005; 
Mieske et al. 2006).  

From the observational side, it is relatively well established that 
the red massive early-type galaxies have at least some lighter elements 
enhanced relative to Fe-peak elements by about 0.3 - 0.4 dex (e.g., 
Worthey et al. 1992; Lee \& Worthey 2005).  This abundance pattern 
may resemble that of halo $\alpha$-element enhancement, although this 
has not been thoroughly proven \citep{wor98}.  
There are several theoretical spectrophotometric studies that consider 
$\alpha$-enhancement in order to address those observations (e.g., 
Thomas et al. 2003; Lee \& Worthey 2005; Coelho et al. 2007; 
Schiavon 2007; Lee et al. 2009).  However, only solar-scaled SBF 
model predictions have been calculated in the past (\citealt{wor93, 
wor94, liu00, bla01, lee01, mei01, can03, mou05, rai05, ma06}).

Moreover, because of the nature of the SBF method, which is far 
more sensitive to the brighter stars compared to the integrated 
photometry, thermally pulsing-asymptotic giant branch (TP-AGB) phase 
is predominately important (e.g., Liu et al. 2000; Mouhcine et al. 2005; 
Raimondo et al. 2005).  The TP-AGB phase is the last stage of AGB 
evolution.  After the high-mass main-sequence (MS) stars and 
supergiants fade away, it is AGB stars that dominate the integrated 
bolometric light until full-fledged red giant branch (RGB) stars are 
looming.  After t $>$ 2 Gyr, the RGB tip becomes nearly as bright 
as the AGB tip but is much more numerously populated, by nearly a 
factor of 10 (e.g., Ferraro et al. 2004; Mucciarelli et al. 2009).  
The recent development of detailed 
studies of TP-AGBs (e.g., Maraston 2005; Maraston et al. 2006; 
Lee et al. 2007b; Marigo et al. 2008) can be thoroughly inspected with 
observations by the SBF method.  In this study, we explore the question 
of how the $\alpha$-elements (e.g., O, Mg, Si, S, Ca, Ti) and 
TP-AGBs affect the galaxy colors and SBF magnitudes.

\section{Models}

The present stellar population synthesis models are based upon the 
Teramo BaSTI isochrones\footnote{http://193.204.1.62/index.html} 
(\citealt{pie04, pie06, cor07}).  
The models we compute are all single starbursts, characterized by a 
single age and metallicity.  In reality, galaxies are generally composed 
of mixtures of stellar populations with a range of ages and metallicities.  

We have, however, deliberately employed a simplified (i.e., single burst) 
star formation prescription in this study so that we can explore 
the importance of varying one parameter, namely the $\alpha$-element 
enhancement.  In this regard, our single-burst models approximately 
represent the luminosity weighted mean age and metallicity for the stellar 
systems in question.  Following our previous models (Lee et al. 2007b, 2009), 
we employ the standard Salpeter (1955) initial mass function.  The low-mass 
cutoff is 0.5 $M_{\odot}$ as given in the Teramo BaSTI isochrones.  
We do not consider the stochastic nature of the TP-AGB in this study.  
We merely calculate the SBF magnitudes and the broadband colors 
using the isochrones as they are available from stellar modelers 
websites.  It is our intention though for future study to investigate 
those stochastic variations.

In $\S$ 2.2, we elaborate the nature of the $\alpha$-enhancement 
in stellar models, particularly the different definitions and degrees of 
$\alpha$-element mixture by different groups and their implications.

\subsection{HRDs and CMDs}

Here we present, for the first time,  $\alpha$-element enhanced 
SBF models and compare them with observations.  Before we fully 
examine the model outputs and the comparisons with observations, 
however, we first look into the $\alpha$-element effects at the 
H-R diagrams (HRDs) and the color-magnitude diagrams (CMDs) as well as 
at the emergent fluxes.  The reason that we primarily employ the Teramo 
BaSTI Isochrones in this study is because the Teramo BaSTI stellar models 
provide both solar-scaled and $\alpha$-enhanced isochrones with correct 
matching of stellar model atmospheres (Cassisi et al. 2004) all the way 
to the full TP-AGB stages that are crucial for the SBF calculations.

Figure 1 contrasts the solar-scaled standard (sss; by ``standard,'' we
indicate that they adopt {\em no} convective-core overshooting) Teramo 
BaSTI isochrones with $\sim$0.4 dex $\alpha$-element enhanced ones (aes) 
in the $log  T_{eff}$ vs.\ $log  L/L_{\odot}$ plane.  At solar 
metallicity ($Z$=0.02), the Teramo BaSTI sss and aes isochrones 
with (right panels) and without (left panels) TP-AGBs are compared 
at three given ages (1, 5, 13 Gyr).  
The sss isochrones are solid lines, while the aes ones are depicted 
with dashed lines.  To guide the eye, RGB tips are denoted with squares 
for the sss and triangles for the aes, respectively.  Bottom panels show 
the details in the giant branches.  In general, the aes isochrones are of 
slightly higher temperature in the red giant branch ($\sim$45K) as well as 
in the upper MS compared to the solar-scaled ones.  They also show 
slightly lower luminosity in the subgiant branch at 1 Gyr.  These effects 
are a reflection of Fe-depression (see Figure 10 of Dotter et al. 2007a) 
traded with $\alpha$-element enhancement at fixed total metallicity, $Z$.  
The right panels of Figure 1 also show that at younger 
ages (t $<$ 5 Gyr), the TP-AGBs go far cooler and brighter compared to 
their RGB tips (e.g., Iban \& Renzini 1983). 

$V - I$ vs.\ $I$ color-magnitude diagrams are displayed in 
Figure 2 in order to compare the sss Teramo BaSTI isochrones with 
the aes ones.  RGB tips are again denoted with squares for the sss 
and triangles for the aes, respectively, to guide the eye.  It is shown 
that the $\alpha$-enhanced red giant branches (RGB), especially at the 
upper part ($I$ $<$ $-$2), are relatively redder and brighter in this 
$V - I$ vs.\ $I$ color-magnitude diagrams compared to the solar-scaled 
ones.  The right panels of Figure 2 illustrate that the $\alpha$-enhanced 
TP-AGBs are also relatively redder and brighter in the $V - I$ vs.\ $I$ 
color-magnitude diagrams compared to the solar-scaled ones.  
This is mainly because of the oxygen-enhancement among 
$\alpha$-elements (see Figure 3).  The next section addresses
this issue in detail.

\subsection{Clarification of $\alpha$-element Enhancement in Stellar Models}

There are now several versions of $\alpha$-element enhanced stellar models in
the literature (e.g., Salasnich et al. 2000; Kim et al. 2002; Vandenberg et
al. 2006; Dotter et al. 2007ab, 2008) besides the Teramo BaSTI ones of
Pietrinferni et al. (2006) that we employ in this study.  One should, however,
carefully examine (1) whether their $\alpha$-element enhancement is defined at
fixed total metallicity, $Z$ or at fixed [Fe/H] and (2) how their
$\alpha$-element mixture is defined for their $\alpha$-element enhanced
stellar models.  In this study, we are contrasting the Teramo BaSTI stellar
models with the Dartmouth ones (Dotter et al. 2008) and duly note that
different $\alpha$-element mixtures by different groups have significant
differences even though they are all commonly referred to as
``$\alpha$-element enhanced stellar models".

We have scrutinized, for example, the differences between Figure 1 of 
this paper and the results shown by Dotter et al. (2007a).  Compared to our 
Figure 1, Figure 11 of Dotter et al. (2007a) shows almost {\em no temperature 
changes} although they are similarly $\alpha$-element enhanced isochrones 
at fixed $Z$.  The culprit is the $\alpha$-element mixture.  
Although the Teramo BaSTI $\alpha$-enhanced stellar models are of 
[$\alpha$/Fe] $\sim$ 0.4 dex, this is an average only.  
The Dartmouth models' $\alpha$-enhanced mix, on the contrary, is a constant 
enhancement of all $\alpha$-elements with respect to solar ratios.  
From inspection of Table 1 of Pietrinferni et al. (2006), their oxygen abundance 
is very high, close to the Dartmouth models' [$\alpha$/Fe] = +0.8 dex value.  
With this information in mind, the behavior in the H-R diagram of 
Teramo BaSTI's $\alpha$-enhanced isochrones makes sense if they are 
compared with Figure 7 (oxygen-enhanced) in Dotter et al. (2007a).  
It is, in fact, the main reason why the New Standard Stellar Population 
Models (NSSPM) project has opened up new windows on element-by-element 
variations in order to decipher the rather cryptic collection of 
$\alpha$-element mixtures (Dotter et al. 2007a; Lee et al. 2009).

It would be useful to compare different sets of $\alpha$-enhanced stellar
population models to see if they predict the same 
effects of $\alpha$-enhancement on SBF.  Unfortunately, 
we do not have the luxury of investigating 
several different versions of $\alpha$-enhanced SBF models at the moment.  
Padova isochrones and stellar evolutionary tracks are perhaps more 
sophisticated at the TP-AGB stages compared to the Teramo BaSTI ones (e.g., 
Marigo et al. 2008\footnote{Table 1 of Marigo et al. (2008) lists the available 
stellar isochrones including the TP-AGB phase.  Marigo et al. (2008) note 
that the TP-AGB of the Teramo BaSTI isochrones by Cordier et al. (2007) was 
computed in a rather crude way, i.e., without considering the third dredge-up 
events (carbon star formation) and hot bottom burning nucleosynthesis.}), 
but the Padova ones do not yet provide the matching $\alpha$-enhanced
stellar evolutionary tracks and isochrones.  Dotter et al. (2008, Dartmouth 
stellar evolutionary models), Kim et al. (2002, Yonsei-Yale models), and 
Vandenberg et al. (2006, Victoria-Regina models) all do 
provide the $\alpha$-enhanced stellar models as well as the 
solar-scaled ones, but they do not provide the matching TP-AGB stages
that are crucial for the SBF models.  These three sets are
of keen interest because they present the $\alpha$-enhanced
stellar models at fixed [Fe/H] instead of at fixed $Z$ as the Teramo BaSTI 
stellar models do.  Regarding the $\alpha$-enhancement at fixed $Z$, 
a depressed Fe abundance preserves the total metallicity.  
For instance, in the Teramo BaSTI stellar models there is about 
0.35 dex [Fe/H] shift at fixed total metallicity, $Z$, between 
the solar-scaled (sss; solid lines) and the $\alpha$-enhanced 
(aes; dashed lines) models.  The case of fixed [Fe/H] is, 
alas, also less than perfectly straightforward.  In that case, 
the enhancement of the $\alpha$-elements increases the overall metallicity, 
and therefore either the abundance of hydrogen or helium (or both) must be 
modified in order to compensate for the increased~$Z$.

Moreover, there is the stellar atmosphere (emergent flux) issue.  
The Teramo BaSTI $\alpha$-enhanced stellar models 
incorporate the matching $\alpha$-enhanced 
stellar model atmospheres self-consistently in order to generate the 
observables (magnitudes and colors) as described in Cassisi et al. (2004).  
All the other $\alpha$-enhanced stellar models, however, employ the solar-scaled 
stellar atmosphere for their calculations of $\alpha$-enhanced model observables.  
It would be useful to have $\alpha$-enhanced models at fixed [Fe/H] from 
the Teramo BaSTI group, as well as at fixed $Z$, so that the effects of 
the $\alpha$-element variation could be seen more directly, instead of 
mixing in the effects of Fe depression for the $\alpha$-enhanced models 
at fixed $Z$ (Salaris et al. 1993).  In the same context, we are also looking 
forward to implementing the Padova $\alpha$-enhanced stellar models with 
matching $\alpha$-enhanced stellar model atmospheres as they become available.  

Figure 3 demonstrates how the enhancement of each element 
(carbon, nitrogen, oxygen, and iron) modifies 
the emergent fluxes at 4000 K and log g = 0.5, which is the typical 
temperature and surface gravity of the upper RGBs and AGBs.  
The lower left panel of Figure 3 
illuminates that the oxygen-enhancement generates a brighter
$I$-band luminosity and a redder $V - I$ color.  
From Table 1, it is noted that the $V - I$ color becomes 0.031 mag 
redder because of the 0.3 dex oxygen-enhancement.  Being a 
dominant $\alpha$-element (e.g., O, Mg, Si, S, Ca, Ti), 
this behavior from the oxygen-enhanced 
spectrum is very useful in order to understand the $V - I$ colors and 
$I$-band magnitudes seen in Figure 2 as well as the model results 
in Figure 4 that we discuss in the following section.  
The upper panels illustrate the carbon- and nitrogen-enhanced 
spectra again at 4000 K and log g = 0.5.  They are, in general, 
displaying the opposite from what we see in lower left panel of 
Figure 3.  By comparing the C-, N-, and O-enhanced spectra, 
it is interesting to find that many features around the $I$-band 
are CN-bands and they are more sensitive to carbon abundance 
than to nitrogen.  The oxygen-enhanced spectrum 
in the lower left panel of Figure 3  illustrates that 
increasing oxygen abundance soaks up more C into the CO molecule, 
decreasing C$_2$, CH, and CN feature strengths.

The lower right panel of Figure 3 shows the iron-enhanced spectrum 
at 4000 K and log g = 0.5.  It is evident that many strong iron 
absorption line features occur around the $UBV$-bandpasses.  
From Table 1, it is noted that the $U - B$ color becomes 0.084 mag 
redder because of the 0.3 dex iron-enhancement.  Moreover, it is 
worthwhile to emphasize that $I$-band luminosity is relatively 
insensitive to the iron-abundance.  The lower right panel of Figure 3 
is quite helpful in understanding the observed
color-magnitude effects because 
the $\alpha$-enhanced Teramo BaSTI stellar models 
are essentially equivalent to Fe-depressed ones.

\section{Results}

Having discussed the nature of the $\alpha$-element enhancement in 
terms of the isochrones and the emergent fluxes, we now present 
the integrated $\alpha$-enhanced SBF models and broadband colors.  
Following our initial results on this topic (Lee et al. 2007a), 
there have been critical updates from the Teramo BaSTI stellar 
models lately.  The $\alpha$-enhanced isochrones and stellar evolutionary 
tracks have been recomputed for $Z$ $>$ 0.001 after 
employing the low-temperature opacities by Ferguson et al. 
(2005)\footnote{See ``News'' in 3/30/2007 and 5/16/2008 at 
http://193.204.1.62/index.html for the details.}.  

Figure 4 shows our new $I$-band SBF model calculations (aes: 
$\alpha$-enhanced standard, sss: solar-scaled standard, where `standard' 
again means {\em no} convective core overshooting) as a function of 
integrated $V - I$  colors based upon the recently 
updated (after May 2008) Teramo BaSTI isochrones.  Two observational 
fiducial lines (thick bent straight lines) are overlaid with our 
theoretical models.  The line on the blue-side is $I$-band 
SBF Mag = $-$2.25 + 2.44 $\times$ [($V - I$) $-$ 1.00] 
from Mieske, Hilker, \& Infante (2006)\footnote{In Mieske, Hilker, \& 
Infante (2006), $I$-band SBF Mag = $-$2.13 ($\pm$0.17) + 
2.44 ($\pm$1.94) $\times$ [($V - I$) $-$ 1.00] for 0.8 $<$ $V - I$ $<$ 1.10.  
We have shifted within the permitted errors in order to match with the 
empirical line by Tonry et al. (2000).} for dwarf blue galaxies, 
while the line on the red-side is $I$-band SBF Mag = $-$1.68 + 
4.5 $\times$ [($V - I$) $-$ 1.15] from Tonry et al. (2000)\footnote{
We have adopted 0.06 mag zero-point correction by 
Blakeslee et al. (2002) to match the final set of $H_0$ 
Key Project Cepheid distances from Freedman et al. (2001).} for massive 
red galaxies.  It is worth mentioning here that we use the observational 
fiducial lines merely as a ``sanity check''.  As we mentioned 
in section 2, we do not consider the stochastic nature of the TP-AGB 
in this study.  Therefore, we want to make it clear that our aim 
in this study is the blunt investigation of the effects of $\alpha$-element 
enhancement and the thermally pulsing-asymptotic giant branch (TP-AGB) 
stars on the surface brightness fluctuation (SBF) magnitudes and broadband 
colors of simple stellar populations.  

Compared to our earlier results (Lee et al. 2007a), the differences between 
the solar-scaled and the $\alpha$-enhanced models are smaller mainly 
because of the Teramo BaSTI isochrones updates (due to the 
low temperature opacities by Ferguson et al. 2005 for the $\alpha$-enhanced 
stellar models).  At solar metallicity ($Z$=0.02, filled squares), it is found 
from Figure 4 (also from Table 3) that the $\alpha$-enhanced models become 
about 0.02 mag redder and 0.35 mag brighter 
in this integrated $V - I$  vs.\ $I$-band SBF magnitude plane compared to the 
solar-scaled ones mostly because of the oxygen-enhancement as we noted in 
Figures 2 and 3.  The right panel of Figure 4 displays 
the integrated $V - I$ colors vs.\ $I$-band SBF 
magnitudes when the TP-AGBs are not included in the calculations.  
The general trend of the $\alpha$-enhanced models at solar metallicity 
($Z$=0.02, filled squares) becoming redder and brighter compared to 
the solar-scaled ones for $t\geq1$ Gyr is mostly intact.  It is 
seen, however, that the $I$-band SBF magnitudes are much too faint 
without the TP-AGBs to match the observations, especially at 
the metal-poor end.  

Figure 5 is similar to Figure 2, but here displays the comparison of the sss 
and the aes Teramo BaSTI isochrones in $V - I$ vs.\ $I$ color-magnitude diagrams 
at $Z$=0.0003.  It is important to note here that compared to Figure 2, the 
$V - I$ colors and the $I$-band magnitudes are hardly changed with the 
$\alpha$-enhancement on this very metal-poor side even at the 
upper RGB.  It explains the comparably smaller effects of $\alpha$-enhancement 
on SBF models and integrated broadband colors at $Z$=0.0003 in Figure 4.  
It is also interesting to see from the left panels (upper: without post-RGB, 
lower: with post-RGB but without TP-AGB) that the blue horizontal-branch 
of 13 Gyr overlaps in $V - I$ color and $I$-band magnitude 
with the MS turnoff of 1 Gyr here at $Z$=0.0003.  
In right panel of Figure 5, TP-AGBs are additionally depicted with 
thicker lines.  It is seen that the TP-AGBs are more than 1 mag redder 
in $V - I$ and 1 - 2 mag {\em brighter} in $I$-band magnitude 
compared to their RGB tips, particularly on this very 
metal-poor side (cf., right panels of Figure 2).  It explains the importance 
of the TP-AGBs on SBF models and integrated broadband colors 
at $Z$=0.0003 in Figure 4 (see also Figure 14).

We also address the convective core overshooting issue at young 
ages (t $<$ 5 Gyr).  Figure 6 is similar to the left panel of Figure 4, 
but here integrated $V - I$ colors and $I$-band SBF 
magnitudes using the Teramo BaSTI aes (dashed lines; $\alpha$-enhanced and
without convective core overshooting) and aeo (solid lines; $\alpha$-enhanced
and with convective core overshooting) isochrones are compared at three 
different relatively young ages.  Note that at 1 Gyr and solar metallicity
(filled squares) overshooting effects make the integrated $V - I$ colors 
0.05 mag bluer and the $I$-band SBF magnitudes 0.18 mag fainter 
as indicated with an arrow (see also Table 3).  
At 5 Gyr, however, 
it is seen that the overshooting effects become negligible.  Figure 7 is 
similar to the right panels of Figures 1 and 2, but the aes (dashed lines) 
and the aeo (solid lines) Teramo BaSTI isochrones are compared in the H-R 
diagrams and the C-M diagrams at 1 and 5 Gyr for $Z$=0.02.
The RGB tips are denoted with triangles for the aes and circles 
for the aeo, respectively, to guide the eye.  It is interesting to note 
that the aeo models (1) have a hotter upper MS near the turnoff 
compared to the aes ones and (2) do not go all the way 
to the RGB tip at 1 Gyr (see also Lee et al. 2007b; Yi 2003).  
However, the aes and aeo models become virtually identical by 5 Gyr.  
The behavior due to the overshooting effects on the integrated $V - I$ 
colors (bluer because of the bluer upper MS 
near the turnoff and the fainter RGB) and on the $I$-band SBF magnitudes 
(fainter because of the less developed upper RGB) that we 
described in Figure 6 can be understood from this C-M diagram.

\subsection{Comparison with Earlier Models}

Figures 8 and 9 compare our scaled-solar (sss) models with other recent 
models available in the literature.  Figure 8 contrasts our models 
based on the Teramo BaSTI `sss' isochrones with Raimondo et al. (2005), 
while Figure 9 does that with Mar\'in-Franch \& Aparicio (2006)\footnote{
Among three different models in Mar\'in-Franch \& Aparicio (2006), we show 
the one with Pietrinferni et al. (2004) isochrones.}.  The large filled 
symbols are used to indicate the solar metallicity in order to guide the eye.  
Both Figures 8 and 9 demonstrate that there are significant differences 
among models mostly because of the different treatment of the TP-AGB stars.  
Hence, we have collected several other available models 
in the literature and listed them in Table 2.  We have only listed their 
integrated $V - I$ colors and $I$-band SBF magnitudes at 5 and 13 Gyr at 
solar metallicity.  


Table 2 tells that there are non-negligible disparities 
among models {\em at the same age and metallicity}.  
The two different Worthey (1994) models\footnote{
http://astro.wsu.edu/worthey/dial/dial$\_$a$\_$model.html} 
evidently show that the different input ingredients (i.e., isochrones) 
make significant ($\sim$0.1 mag in $V - I$ and $\sim$0.55 mag in $I$-band 
SBF magnitudes) disparities {\em at the same age and metallicity}.  
Obviously, depending upon the ingredients and recipe the modelers 
adopt such as (1) isochrones, particularly the late-evolutionary 
stage evolutionary tracks such as RGB, AGB, and TP-AGB, (2) mass-loss 
scheme such as $\eta$ in Reimers (1975), (3) stellar library to convert 
the temperature and luminosity to colors and magnitudes, and (4) the IMF, 
one can get significantly different model outputs.  It is therefore 
considerably important to carefully study the ingredients and recipe of 
different models before use.  A rigorous test of the 
integrated photometric models is imperative.

On a positive note, however, our models agree extremely well with 
other very recent works using {\em independent} stellar population models.  
For example, our scaled-solar (sss) models and 
Percival et al. (2009; P09) that are based upon the most up-to-date 
Teramo BaSTI with $\eta$ = 0.4 and Salpeter IMF agree each other 
within 0.003 mag in $V - I$.  Moreover, our scaled-solar models that are 
based upon the most up-to-date Padova isochrones with Salpeter IMF and 
Padova SSP models\footnote{They are available from 
http://stev.oapd.inaf.it/cgi-bin/cmd} (Marigo et al. 2008) agree 
each other within 0.002 mag in $V - I$ at solar metallicty at 13 Gyr.  
Our models based upon the different isochrones will be compared with 
one another in detail in the following section.

\subsection{Comparison of Padova and Teramo/BaSTI Solar-Scaled SBF Models}  

Having discovered the non-negligible impacts of the input ingredients 
(i.e., isochrones) on the SBF magnitudes and broadband colors, here we 
calculate those quantities by employing stellar models from different 
groups but at the same {\em solar-scaled} composition.  
In this study, we contrast the widely-used two stellar models, 
the Padova\footnote{http://stev.oapd.inaf.it/cgi-bin/cmd} 
and the Teramo BaSTI stellar models\footnote{http://193.204.1.62/index.html} 
at the same {\em solar-scaled} chemical composition.  It is worthwhile 
to reiterate that $\alpha$-enhanced stellar models could be even more 
diverse than expected because of differing definitions of ``alpha-element''.

Figure 10 is similar to Figure 4, but here we contrast the $I$-band 
SBF models as a function of integrated $V - I$ colors at given ages 
and metallicities using two different stellar models but at the same 
{\em solar-scaled} chemical composition {\em with convective core 
overshooting}.  One is using the Teramo BaSTI (solid lines) and 
the other is using the Padova (dashed lines) stellar models.  
We employ their latest stellar models, which we directly download from 
their websites (see footnotes \#1 and \#9, respectively) in order to
calculate the SBF predictions.  We have employed the `sso' (solar-scaled 
with convective core overshooting) Teramo BaSTI models because we note 
that the Padova stellar models employ the convective core overshoot 
as the default in their models\footnote{Please refer to footnotes 
\#1 and \#9 for the detailed treatment of the convective core 
overshooting from each stellar model group.}.  
To guide the eye, solar metallicity models of varying ages are 
marked with filled squares.  The two sets of stellar models result in 
significant disparities.  In general, integrated $V - I$ colors 
based on the Padova isochrones are comparatively redder than that from the 
Teramo BaSTI (see Table 3).  It is 
also noted from the left panel that the $I$-band 
SBF magnitudes based upon the Teramo BaSTI become much fainter 
($>$ 1 mag) at younger ages (t $<$ 5 Gyr) with $Z$ $\geq$ 0.0004 
compared to Padova models.  In order to ascertain whether the remarkable 
dissimilarities are mainly caused by the rather poorly understood bright 
TP-AGB stars, we display the same models without TP-AGBs in the right 
panel of Figure 10.  The differences using two different 
stellar models persist even without the TP-AGBs although they are 
much less compared to that with the TP-AGBs.  

From Figure 10, however, it is evident that the inclusion of 
TP-AGB stages is indeed necessary in matching the observations, which 
are represented by the thick bent lines.  The systematic redder 
integrated $V - I$ colors using the Padova stellar models compared to 
that using the Teramo BaSTI ones can be understood from the fact
that the Padova RGBs are systematically cooler and redder than the 
Teramo BaSTI ones, as illustrated in Figures 11 and 12.  
For a clearer understanding of the cause of the dissimilarities of 
the stellar population model $I$-band SBF predictions as well as 
the integrated $V - I$ colors using the the Padova and the Teramo BaSTI 
stellar models, we illustrate the comparison of the Padova and the 
Teramo BaSTI stellar models in the H-R diagrams and the C-M diagrams 
in the following figures.  

Figure 11 shows the comparison of the Teramo BaSTI (solid lines) 
and the Padova (dashed lines) isochrones at the same {\em solar-scaled} 
chemical composition in the HR diagrams without (left panels) 
and with (right panels) TP-AGB stages, 
respectively.  At solar metallicity, three ages (1, 5, 13 Gyr) are compared.  
RGB tips are denoted with squares for Teramo BaSTI and triangles for 
Padova, respectively.  From Figure 11, it is noted that the MSTO and 
RGB temperatures of the older ages (t $\geq$ 5 Gyr) are generally cooler 
in the Padova stellar models, which cause the redder integrated $V - I$ 
colors compared to those using the Teramo BaSTI ones in Figure 10.  
The small RGB temperature differences may be ascribed to the uncertainties 
of the convection treatment in the RGB stars.  From the right panels of 
Figure 11, it is also noted that at younger ages (t $<$ 5 Gyr), 
the TP-AGBs go far cooler and brighter than their RGB tips.  

Figure 12 displays the comparison of the Teramo BaSTI (solid lines) and 
the Padova (dashed lines) isochrones in $V - I$ vs.\ $I$ color-magnitude 
diagrams at the same {\em solar-scaled} composition.  Symbols for the RGB 
tips are same as in Figure 11.  It is noted that the Padova RGBs, especially 
at the older ages (t $\geq$ 5 Gyr), are relatively redder than the Teramo 
BaSTI ones in these C-M diagrams.  From the right panels of Figure 12, 
it is seen that, in general, the 
Padova TP-AGBs models are more complex than the Teramo BaSTI TP-AGBs.  
It is further noted from the right panels of Figure 12 that there are 
discontinuities for the Padova stellar models at the onset of the TP-AGB 
stage caused by structural changes (Marigo et al. 2008).  For instance, 
the H-exhausted core mass on the TP-AGB starts to increase (and is set to 
zero before the TP-AGB).

\subsection{Near-IR SBF Models}  

Having found that (1) there are significant differences in the $I$-band SBF 
model predictions using the Padova and the Teramo BaSTI models and (2) the 
inclusion of TP-AGBs is crucial to match the observations, we now 
extend our investigation to longer wavelengths, including near-IR SBF models 
where the effects of the TP-AGB stage are considerably more pronounced.

Figure 13 is similar to the left panel of Figure 10, but here we contrast 
the $z_{850}$-band SBF models based on the Padova stellar models\footnote{
The reason that we show the $z_{850}$-band SBF models only based on the 
Padova stellar models is because the $z_{850}$-band magnitude is not yet 
available from the BaSTI websites.} as a function of integrated 
$g_{475} - z_{850}$ colors at given ages and metallicities.  The thick 
curved line is an empirical relation from Blakeslee et al. (2009).  
The {\em HST} ACS/WFC photometric systems\footnote{The F475W and F850LP 
are equivalent to $g_{475}$ and $z_{850}$-bands, respectively.} that we 
present here are all AB magnitudes in order to be consistent with the 
observations.  We have converted the Vega magnitude system to AB magnitude 
system by employing F475W (AB) = F475W (Vega) $-$ 0.101 and F850LP (AB) = 
F850LP (Vega) + 0.565 from Sirianni et al. (2005).  To guide the eye, 
solar metallicities are marked with filled squares and 8 and 13 Gyr are 
displayed in different colors.  

The arrows in Figure 13 at 5, 8, and 13 Gyr indicate the estimated effects 
on the models from 0.4 dex $\alpha$-enhancement.  Since the Padova group 
do not yet provide the matching $\alpha$-enhanced stellar models, 
those arrows are inferred from the $\alpha$-enhanced $I$-band SBF and 
integrated $V - I$ model predictions based on the Teramo BaSTI stellar 
models.  We have used the 
($V - I$) = 0.603 $\times$ ($g_{475} - z_{850}$) + 0.375 and $I$-band SBF mag 
= 0.930 $\times$ $z_{850}$-band SBF mag + 0.433.  It is interesting to find 
that the effects of the $\alpha$-enhancement on the $z_{850}$-band SBF models 
alleviate the mismatches with the observations on the red-side, 
$g_{475} - z_{850}$ $>$ 1.3.   
The mismatch that is seen on the blue-side in Figure 13, 
$g_{475} - z_{850}$ $<$ 1.1 
may be relieved if we could employ the BaSTI isochrones as we saw 
in Figure 10 that the BaSTI isochrones generate systematically much  fainter 
SBF magnitudes compared to Padova isochrones at younger ages (t $<$ 5 Gyr) 
with $Z$ $\geq$ 0.004.  Biscardi et al.\ (2008) provide another 
interesting comparison of $z_{850}$-band SBF observations with model 
predictions, but using the Teramo SPoT models\footnote{
http://193.204.1.79:21075/models.html}. 

Figure 14 contrasts the {\em HST} NICMOS F160W-band SBF models as a function 
of integrated $V - I$ colors using the Padova and the Teramo BaSTI stellar 
models at the same {\em solar-scaled} chemical composition.  The thicker 
line is the empirical sequence from Jensen et al. (2003).  The {\em HST} 
NICMOS photometric bandpass, F160W, that we present here is Vega magnitudes 
in order to be consistent with the observations.  We have converted 
$JHK$-band SBF models using the Teramo BaSTI isochrones to F160W-band SBF 
ones by employing equation (3) of Jensen et al. (2003).  As expected, the 
importance of the inclusion of the TP-AGB stage is considerably greater here 
in the near-IR compared to Figure 10.  For instance, at 1 Gyr and 
solar metallicity, the F160W-band SBF magnitude predictions from the Teramo 
BaSTI models become $\sim$1.8 mag fainter without TP-AGBs (see Table 3).  
Also, it is noted that the trend between F160W-band SBF models and 
the integrated $V - I$ colors reverses, in the sense that  
the near-IR SBF magnitudes become brighter with bluer colors 
when the TP-AGB is included, but fainter at bluer colors when it is omitted.
The significant differences between the models based on the 
Teramo BaSTI and the Padova models are still noticeable even without TP-AGBs 
from the right panel of Figure 14 at the same {\em solar-scaled} chemical 
composition.   

Figure 15 is similar to the left panel of Figure 14, but displays 
the $\alpha$-element enhancement effects for the {\em HST} 
NICMOS F160W-band SBF models based on the Teramo BaSTI.  The thick 
observational fiducial line is the same as in Figure 14.  
The F160W-band SBF magnitude differences between the 
solar-scaled and the $\alpha$-enhanced models using the Teramo BaSTI 
stellar models are relatively small, less than 0.2 mag (see Table 3).  
The reason for the mismatches with the observations 
at the very red end remains 
to be resolved.  Compared to the observations, the left panel of Figure 4 
showed that the $I$-band SBF models are a bit fainter, but here in 
Figure 15, F160W-band SBF models are brighter.  Figure 1 of 
Blakeslee et al. (2009), however, shows that the very red galaxies 
become progressively fainter in $z$-band SBF causing the non-linear 
relation between SBF magnitudes and the integrated colors as we have seen 
in Figure 13.

\section{Summary and Discussion}  

We have presented for the first time the effects of $\alpha$-element
enhancement on SBF models and have compared these models with observations.  
For this purpose, we have employed the Teramo BaSTI Isochrones 
in this study.  In general, we find that the $\alpha$-element enhanced 
$I$-band SBF magnitudes are about 0.35 mag brighter and 
their integrated $V - I$ colors are about 0.02 mag redder 
mostly because of the oxygen enhancement effects on the upper 
RGBs and AGBs.  Moreover, the importance of the TP-AGB stages for the 
$I$-band and near-IR band SBF magnitudes is illustrated and it becomes 
clear that TP-AGBs are the indispensable component in order to 
match the theoretical predictions with the observations.  It is also noted 
that the TP-AGBs are more important in the metal-poor regime 
as shown in Figures 4, 5, 10, and 14.  

We have further shown that the disparity in the SBF model and integrated 
$V - I$ color predictions is non-negligible when different stellar models 
are employed as input ingredients even at the same {\em solar-scaled} 
chemical composition.  It is our understanding that what causes the 
disagreement in the integrated $V - I$ colors and $I$-band SBF models 
using different stellar models are (1) giant branch temperature differences, 
which occur depending upon the treatment of convection and (2) dissimilar 
stellar model atmospheres that are employed in order to convert from theoretical 
planes to observables.  Further comparison of SBF model predictions with 
multi-band observations can help illuminate many of the remaining problems 
in the evolution of bright stars relevant to population synthesis modeling.  

The $I$-band SBF models explored here are 
slightly fainter than the observations at the red end ($V - I$ $>$ 1.15), 
as shown in Figure 4.  A partial solution may come from the non-linear 
behavior between galaxy colors in $g_{475} - z_{850}$ and the $z_{850}$-band 
SBF magnitudes as illuminated by Blakeslee et al. (2009) in the sense that 
the red-end galaxies become rapidly fainter in $z_{850}$-band SBF magnitudes.  
In this context, the $\alpha$-enhanced SBF models help to reconcile the 
disagreement by making the theoretical $I$-band and $z$-band SBF brighter.  
Also, rigorous statistical investigations of model isochrones with TP-AGBs 
that are poorly populated yet very salient because of their prominent 
luminosity should help alleviate the discrepancies between the observations 
and the theoretical predictions as suggested by Cervi\~no et al. (2008), 
Raimondo (2009), and Gonzalez-Lopezlira et al. (2009).  
It is evident from the smaller scatter in the observations compared to 
the theoretical predictions that there is generally less variation 
among the AGB phases in actual galaxies than may be expected based on 
the models.  The observed relations between SBF magnitudes and integrated 
colors are very tight, at least for evolved galaxies.  The future 
sophisticated $\alpha$-enhanced SBF studies in various bandpasses 
should provide some additional constraints on the fine details 
of the calibration.

\acknowledgements 
We thank Michele Cantiello, Santi Cassisi, Aaron Dotter, Leo Girardi, 
Paola Marigo, Reynier Peletier, and Gabriella Raimondo for very insightful 
discussions.  We are also grateful to the anonymous referee for her/his 
careful reading and thoughtful comments that have improved our presentation.  
Support for this work was provided by the NSF through grant AST-0307487, 
the New Standard Stellar Population Models (NSSPM) project and 
by the NASA through grant HST-GO-11083.

\begin{deluxetable}{lrrrrrrrrr}
\tablecolumns{10}
\tabletypesize{\small}
\tablecaption{Color Behavior with Elemental Enhancements for a Star with $T_{\rm eff}=4000$ K, $\log g = 0.5$}
\tablehead{
\colhead{Color} & \colhead{C} 
& \colhead{N}  & \colhead{O} 
& \colhead{Mg} & \colhead{Si}
& \colhead{S}  & \colhead{Ca} & \colhead{Ti} & \colhead{Fe} 
\\  }

\startdata

$U-B$ &    30 &   26 &  -21 &   34 &  -43 &    0 &   16 &   24 &   84 \\
$B-V$ &    16 &   -2 &  -13 &  -35 &   18 &    0 &    7 &   10 &   13 \\
$V-I$ &   -30 &  -28 &   31 &   12 &    0 &    0 &   -4 &    2 &   16 \\

\enddata
\tablecomments{1. All elements scaled individually by +0.3 dex, 
                  except C, which is increased by +0.15 dex.  
               2. Numbers are in milli-magnitude.}
\end{deluxetable}

\begin{deluxetable}{lrrrr}
\tablecolumns{5}
\tabletypesize{\footnotesize}
\tablecaption{Integrated $V - I$ and $I$-band SBF Mag from Different SSP Models}
\tablehead{
\colhead{Models} & \colhead{$V - I$ (5 Gyr)} 
& \colhead{$V - I$ (13 Gyr)}  & \colhead{$\bar {M}_I$ (5 Gyr)} 
& \colhead{$\bar {M}_I$ (13 Gyr)}
\\  }

\startdata

W94, Sal        &  1.186  &   1.312  &  -1.680  &   -1.255 \\
W94, Sal, B94   &  1.148  &   1.210  &  -1.129  &   -0.699 \\
B01, Sal, G00   &  1.140  &   1.240  &  -1.410  &   -0.960 \\
BC03, Cha, B94  &  1.124  &   1.227  & \nodata  &  \nodata \\
BC03, Sal, B94  &  1.141  &   1.255  & \nodata  &  \nodata \\
M05, Kro        &  1.088  &   1.176  & \nodata  &  \nodata \\
M05, Sal        &  1.099  &   1.198  & \nodata  &  \nodata \\
R05, Sca, P04   &  1.174  &   1.234  &  -1.548  &   -1.225 \\
MA06, Kro, B94  &  1.167  &   1.282  &  -1.097  &   -0.664 \\
MA06, Kro, G00  &  1.163  &   1.278  &  -1.290  &   -0.909 \\
MA06, Kro, P04  &  1.124  &   1.243  &  -1.663  &   -1.104 \\
Pad$\_$SSP, Cha &  1.162  &   1.238  & \nodata  &  \nodata \\
Pad$\_$SSP, Sal &  1.172  &   1.261  & \nodata  &  \nodata \\
P09, sss, 0.2   &  1.092  &   1.185  & \nodata  &  \nodata \\ 
P09, sss, 0.4   &  1.089  &   1.177  & \nodata  &  \nodata \\ 
P09, aes, 0.2   &  1.121  &   1.203  & \nodata  &  \nodata \\
P09, aes, 0.4   &  1.117  &   1.193  & \nodata  &  \nodata \\ 
This Work, sss  &  1.086  &   1.178  &  -1.235  &   -0.853 \\
This Work, aes  &  1.111  &   1.192  &  -1.567  &   -1.217 \\
This Work, Pad  &  1.173  &   1.263  &  -1.373  &   -0.831 \\

\enddata
\tablecomments{1. W94 (Worthey 1994) is from 
                  http://astro.wsu.edu/worthey/dial/dial$\_$a$\_$model.html  
                  `Sal' is Salpeter IMF and `B94' is Bertelli et al. (1994).  
               2. B01 (Blakeslee, Vazdekis, \& Ajhar 2001) are 5 and 12.6 Gyr 
                  values from their Table 2.  
               3. BC03 (Bruzual \& Charlot 2003) is from 
                  http://www2.iap.fr/users/charlot/bc2003/  
                  `Cha' is Chabrier (2003) IMF.  
               4. M05 (Maraston 2005) is from 
                  http://www.icg.port.ac.uk/$\sim$maraston/SSPn/colors/SSPcolours$\_$Mar05$\_$JohnsonCousins.tab  `Kro' is Kroupa (2001) IMF.
               5. R05 (Raimondo et al. 2005) is from 
                  http://193.204.1.79:21075/models.html  
                  `Sca' is Scalo (1998) IMF and `P04' is Pietrinferni et al. (2004). 
               6. MA06 (Mar\'in-Franch \& Aparicio 2006) used Kroupa IMF 
                  (Kroupa et al. 2003).  
               7. Pad$\_$SSP (Marigo et al. 2008) is from 
                  http://stev.oapd.inaf.it/cgi-bin/cmd  
               8. P09 (Percival et al. 2009) is from 
                  http://193.204.1.62/index.html  
                  `sss' is scaled-solar and `aes' is $\alpha$-enhanced.
                  Two mass-loss schemes are used; one with $\eta$ = 0.2 
                  and the other with $\eta$ = 0.4.  
               9. This Work, sss and aes adopt the $\eta$ = 0.4.  
                  This Work, Pad is the SSP based on Marigo et al. (2008).}
\end{deluxetable}

\begin{deluxetable}{lrrr}
\tablecolumns{4}
\tabletypesize{\small}
\tablecaption{SBF Magnitude And Integrated $V-I$ Color Differences At $Z$ = 0.02}
\tablehead{
\colhead{Age} & \colhead{$\Delta$ $I$-band SBF} & \colhead{$\Delta$ $F160W$-band SBF} 
& \colhead{$\Delta$ $V-I$} \\
\colhead{(Gyr)} & \colhead{(mag)} & \colhead{(mag)} & \colhead{(mag)} }

\startdata

\sidehead{Effects of 0.4 dex $\alpha$-enhancement using Teramo BaSTI: aes $-$ sss} 

0.5 & -0.380 &  0.168 &  0.058 \\
1   & -0.242 &  0.152 &  0.029 \\
5   & -0.332 &  0.080 &  0.025 \\
13  & -0.364 &  0.066 &  0.014 \\

\sidehead{Effects of TP-AGB (using Teramo BaSTI sss): with TP-AGB $-$ without TP-AGB}

0.5 & -1.084 & -4.571 &  0.068 \\
1   & -0.286 & -1.880 &  0.057 \\
5   & -0.068 & -0.472 &  0.015 \\
13  & -0.046 & -0.207 &  0.011 \\

\sidehead{Effects of convective core overshooting (using Teramo BaSTI): aeo $-$ aes}

0.5 & -0.094 & \nodata & -0.098 \\
1   &  0.180 & \nodata & -0.050 \\
5   & -0.003 & \nodata &  0.004 \\

\sidehead{Effects of stellar models at solar-scaled with TP-AGB: Padova $-$ Teramo BaSTI}

0.5 & -1.444 & -0.736 &  0.041 \\
1   & -0.743 & -0.945 & -0.036 \\
5   & -0.138 & -0.546 &  0.087 \\
13  &  0.022 &  0.006 &  0.085 \\

\enddata
\tablecomments{1. The negative numbers in columns 2 and 3 indicate that the effects make 
               the SBF magnitude brighter and vice versa.
               2. The positive numbers in column 4 indicate that the effects make 
               the integrated $V-I$ color redder and vice versa.}
\end{deluxetable}


\clearpage

\begin{figure}
\epsscale{.8}
\plotone{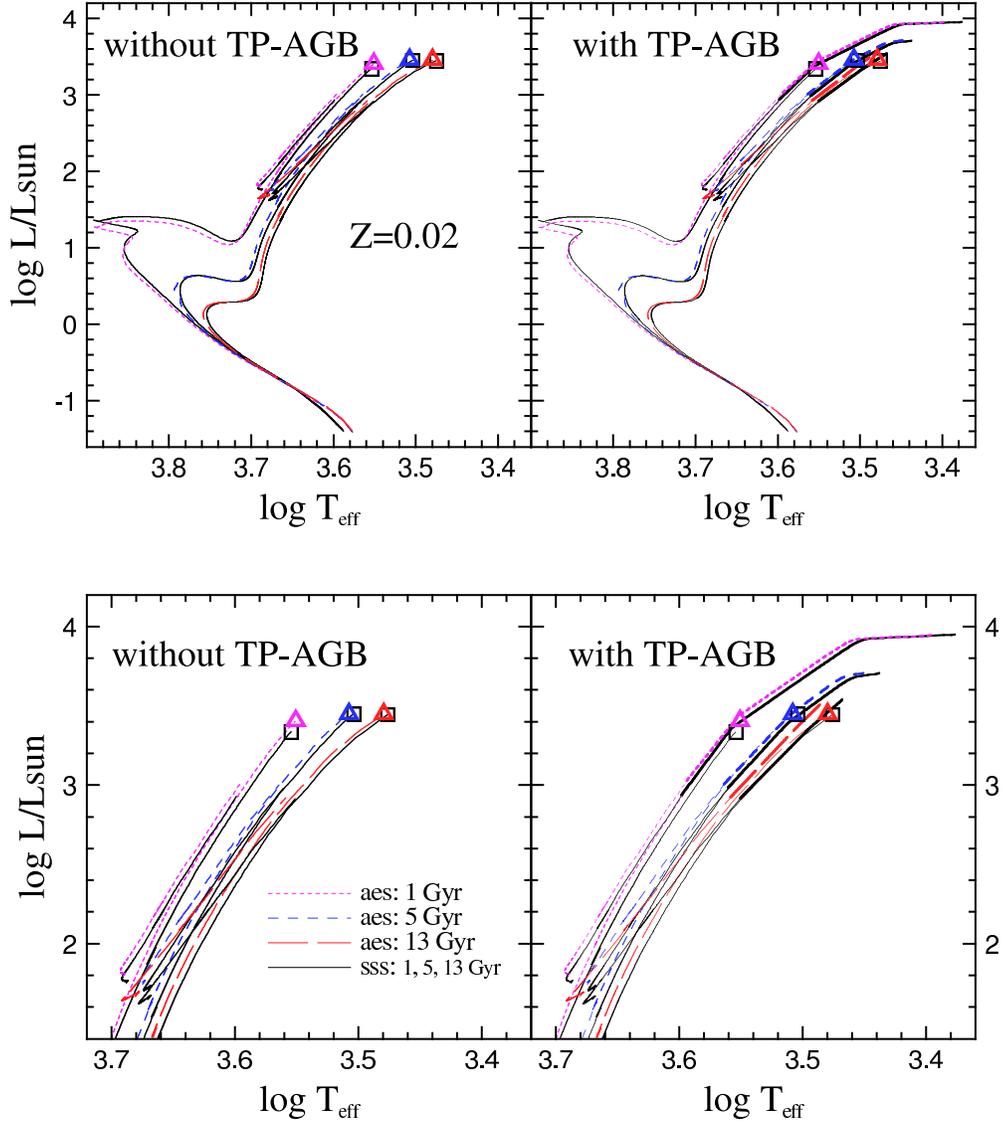}
\caption{Comparison of solar-scaled (sss; solid lines) and 
$\alpha$-enhanced (aes) Teramo BaSTI isochrones in the HR diagram 
without convective core overshooting.  At solar metallicity ($Z$=0.02), 
three ages (1, 5, 13 Gyr) are compared.  Left panels display isochrones 
without thermally pulsing-asymptotic giant branches (TP-AGB), while 
right panels show isochrones with TP-AGBs (thicker lines).  Red giant 
branch tips (RGB tips) are depicted with squares for the sss and 
triangles for the aes, respectively.  Bottom panels show the details 
in the giant branches.  The RGB temperatures are generally slightly 
warmer ($\sim$45K) with $\alpha$-enhancement {\em at fixed total metallicity}, 
partly reflecting the depression of iron abundance.  Note also from the 
right panels that at younger ages (t $<$ 5 Gyr), the TP-AGBs go far 
cooler and brighter compared to their RGB tips.  
\label{fig01}}
\end{figure}

\begin{figure}
\epsscale{.8}
\plotone{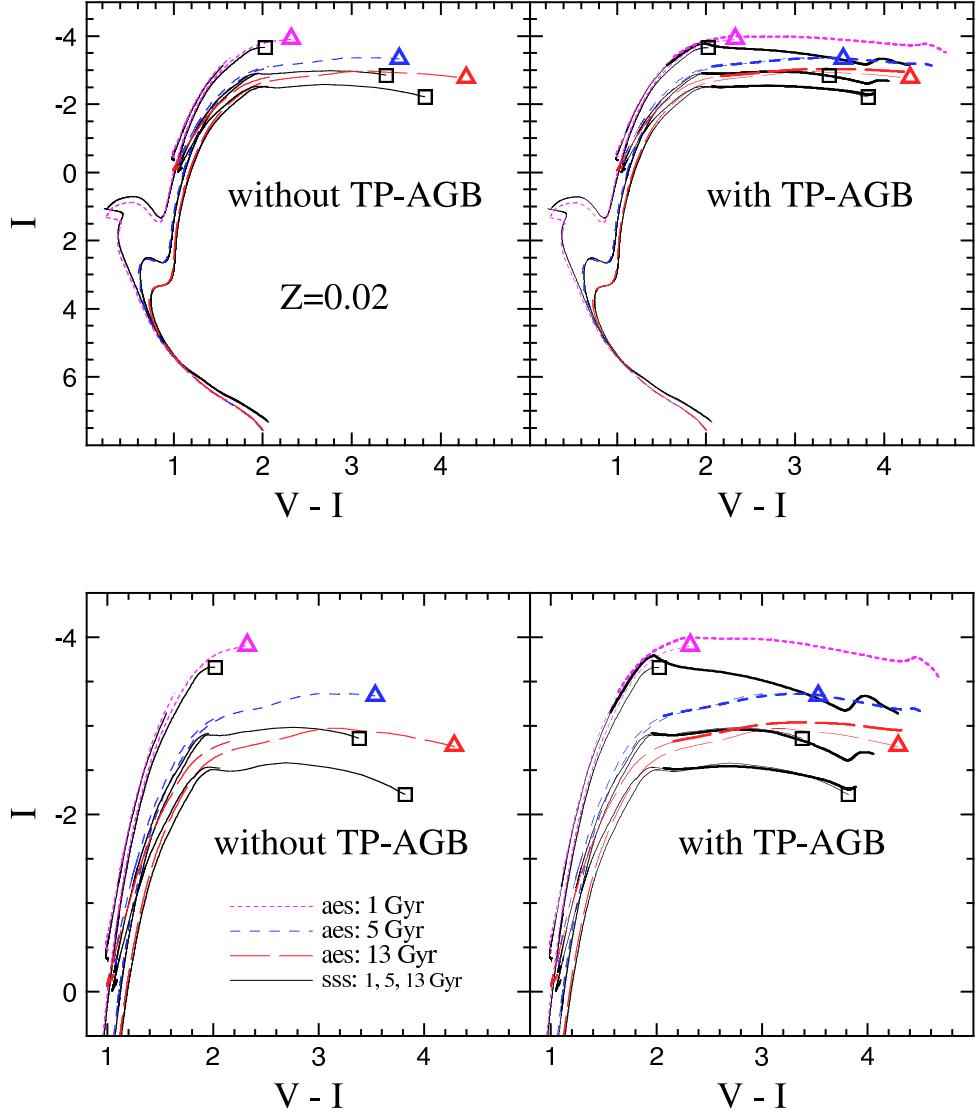}
\caption{Similar to Figure 1, but here we display the comparison of 
the sss and the aes Teramo BaSTI isochrones in the $V - I$ vs.\ $I$ 
color-magnitude (C-M) diagrams.  Symbols for the RGB tips and the TP-AGBs are 
same as in Figure 1.  Note from the left panels that the $\alpha$-enhanced 
red giant branches (RGBs), especially at the upper part ($I$ $<$ $-$2), 
are relatively brighter and redder in this $V - I$ vs.\ $I$ C-M diagrams 
compared to that of the solar-scaled.  Similarly, it is noted from 
the right panels that the $\alpha$-enhanced TP-AGBs are also 
comparatively brighter and redder than the solar-scaled models.  
Moreover, it is seen from the right panels that at younger 
ages (t $<$ 5 Gyr), the TP-AGBs go far redder compared to their RGB tips.  
\label{fig02}}
\end{figure}

\begin{figure}
\epsscale{.8}
\plotone{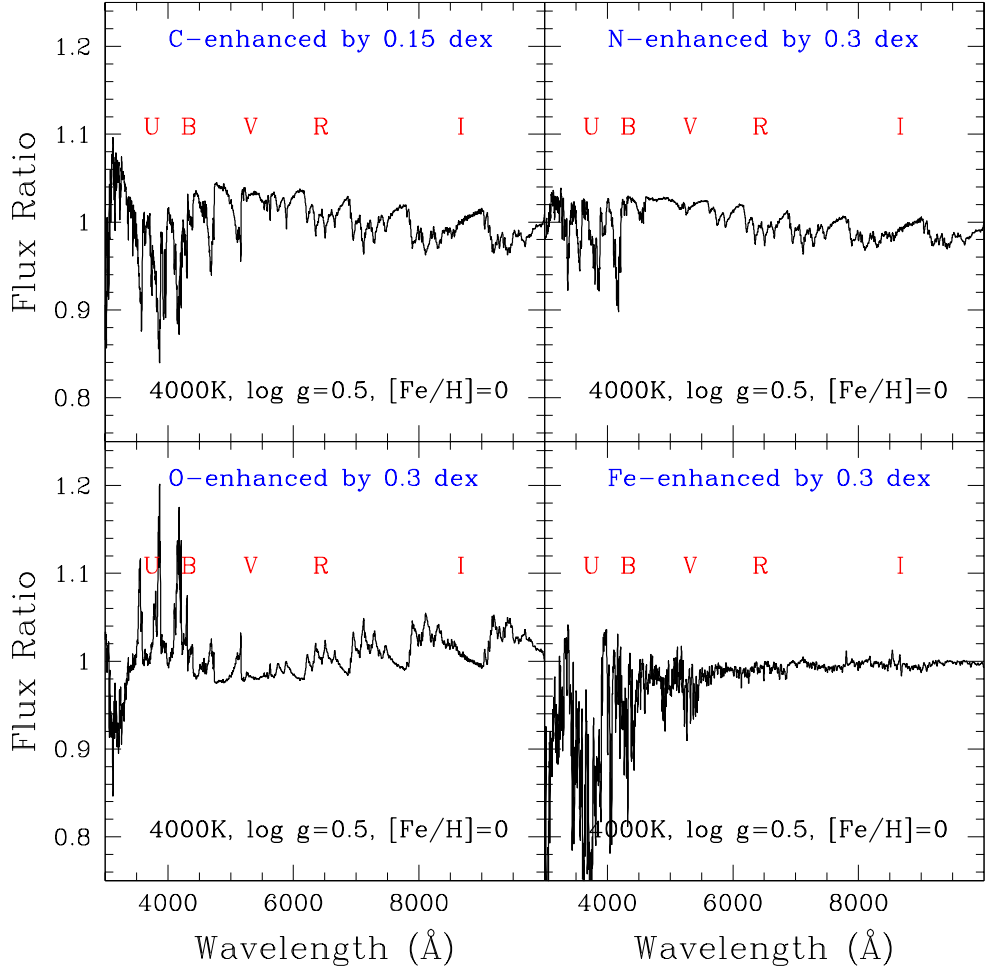}
\caption{At 4000 K, log g = 0.5, and [Fe/H] = 0, the solar-scaled spectrum 
is divided by the 0.15 dex carbon-enhanced spectrum (upper left), 
the 0.3 dex nitrogen-enhanced spectrum (upper right), 
the 0.3 dex oxygen-enhanced spectrum (lower left), and 
the 0.3 dex iron-enhanced spectrum (lower right), respectively.  
$UBVRI$-band filter locations are indicated.  It is seen from the lower left 
panel that the oxygen-enhancement makes the higher 
$I$-band luminosity and the redder $V - I$ color, while 
from the upper panels it is noted that 
carbon- and nitrogen-enhancement make the bluer $V - I$ color 
at this temperature and surface gravity of the typical 
upper RGB and AGB.  From the comparison of C-, N-, O-enhanced cases, 
it is identified that many features around 
the $I$-band are CN-bands and that they are more sensitive to the 
carbon abundance than to the nitrogen.  
Furthermore, it is seen from the lower right panel that many strong iron 
absorption line features are mostly located within the $UBV$-bandpass.  
Note also that the $I$-band luminosity is relatively insensitive 
to the iron-abundance.  
\label{fig03}}
\end{figure}

\begin{figure}
\epsscale{1.}
\plotone{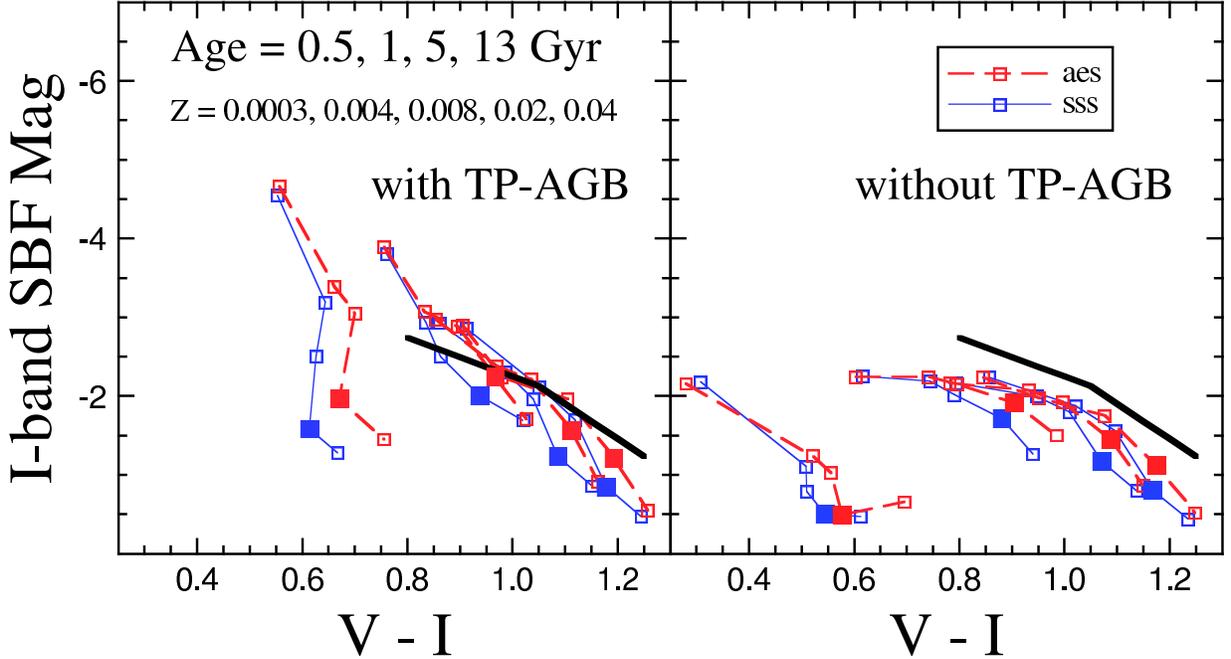}
\caption{$I$-band SBF magnitudes as a function of integrated $V - I$ colors 
are shown at 0.5, 1, 5, and 13 Gyr (left to right) for five different 
metallicities employing the most up-to-date Teramo BaSTI isochrones available 
after May 2008.  Solid lines with squares are solar-scaled models, 
while dashed lines with squares are $\alpha$-enhanced models.  
At given ages, $I$-band SBF magnitudes become fainter and $V - I$ colors 
become redder, in general, with increasing metallicity.  
To guide the eye, solar-metallicity ($Z$=0.02) is depicted with 
filled squares.  Two observational fiducial lines (bent thick solid lines; 
bluer one from Mieske et al. 2006 and redder one from Tonry et al. 2000) 
are compared with our theoretical models.  
The $\alpha$-enhanced models at solar metallicity, 
$Z$=0.02 (filled squares), become redder and brighter (please see 
Table 3 for the values) compared to the 
solar-scaled ones mostly because of the oxygen-enhancement as we noted in 
Figures 2 and 3.  The models without TP-AGBs on the right panel show that 
the general effect of the $\alpha$-enhanced models at solar metallicity 
($Z$=0.02, filled squares) becoming redder and brighter compared to 
the solar-scaled ones for t $\geq$ 1 Gyr are mostly unchanged.  It is 
noted, however, that the $I$-band SBF magnitudes are much too faint 
without TP-AGBs to match the observations.  
\label{fig04}}
\end{figure}

\begin{figure}
\epsscale{.8}
\plotone{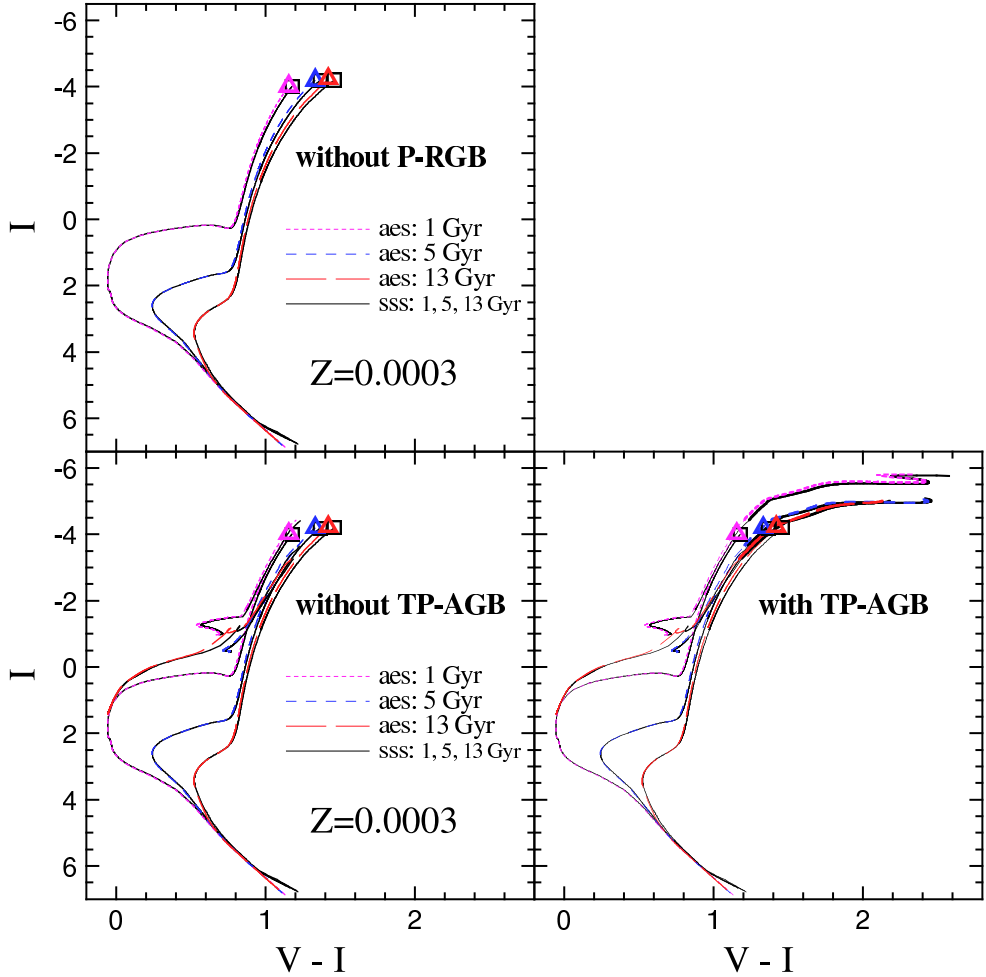}
\caption{Similar to Figure 2, but here we display the comparison of the 
solar-scaled (sss) and the $\alpha$-enhanced (aes) Teramo BaSTI isochrones 
in the $V - I$ vs.\ $I$ color-magnitude diagram at $Z$=0.0003.  
Symbols for the RGB tips and the TP-AGBs are same as in Figure 1.  
Top left panel shows the C-M diagram without post-RGB (p-RGB) stars.  
From left panels, it is interesting to find that the blue horizontal-branch at 
13 Gyr overlaps in $V - I$ color and $I$-band magnitude with the MS 
turnoff of 1 Gyr here at $Z$=0.0003.  Note that compared to Figure 2, the 
$V - I$ colors and the $I$-band magnitudes are hardly changed with the 
$\alpha$-enhancement at this very metal-poor regime even at the 
upper RGB.  It explains the comparably smaller effects of $\alpha$-enhancement 
at $Z$=0.0003 in Figure 4.  Also, note from the right panel that at all 
ages, the TP-AGBs go far redder and brighter compared to their RGB tips, 
particularly at this very metal-poor regime, which explains the importance 
of the TP-AGBs at $Z$=0.0003 shown in Figure 4.
\label{fig05}}
\end{figure}

\begin{figure}
\epsscale{1.}
\plotone{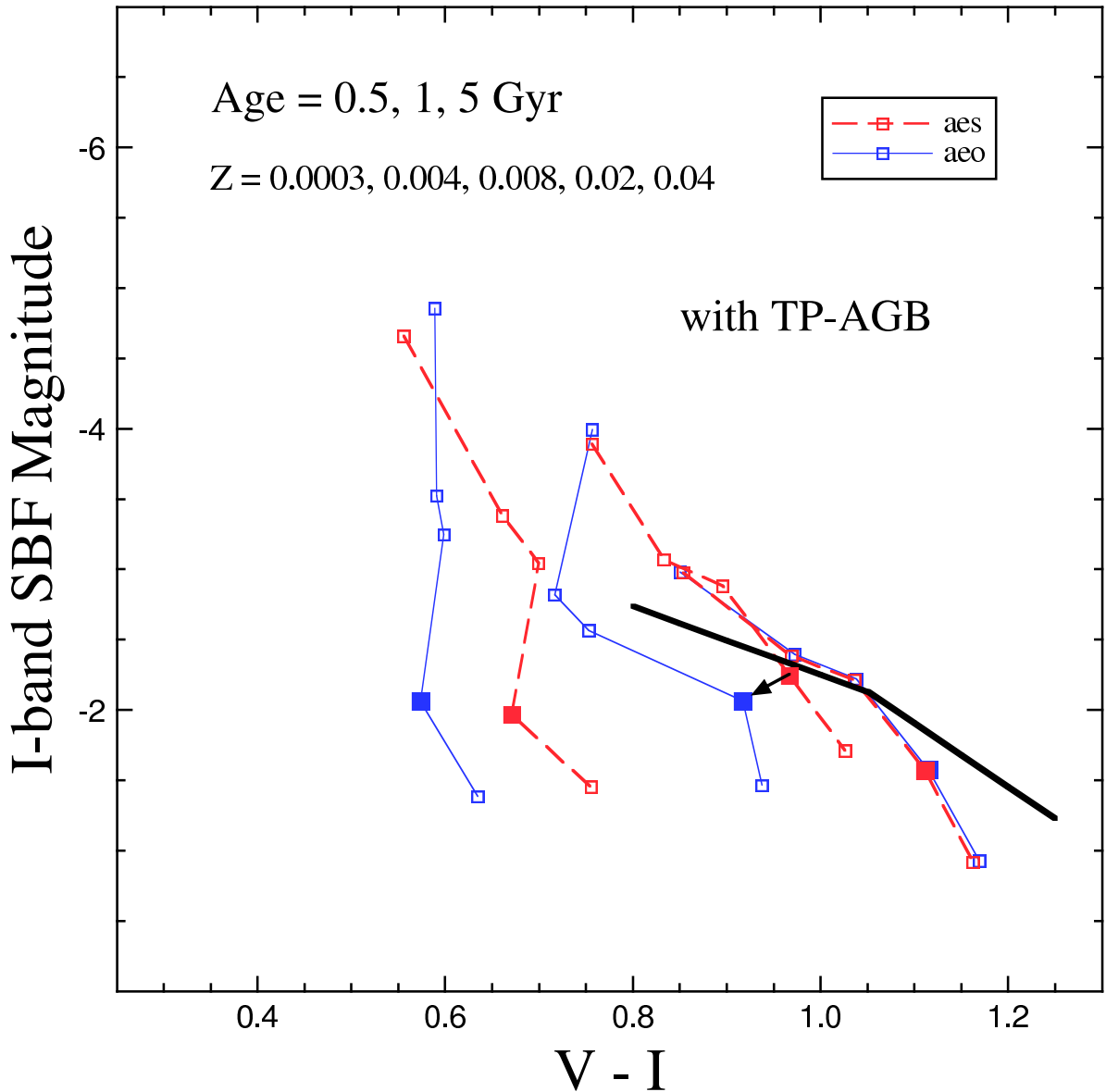}
\caption{Similar to the left panel of Figure 4, but here 
$\alpha$-enhanced $I$-band SBF magnitudes 
as a function of $V - I$ are compared employing the Teramo 
BaSTI aes (dashed lines; without convective core overshooting) and 
aeo (solid lines; with convective core overshooting) isochrones at 
three given ages.  The observational fiducial lines are same as in 
Figure 4.  Note that at 1 Gyr and solar metallicity ($Z$=0.02, filled 
squares), overshooting effects make the integrated $V - I$ colors bluer 
and the $I$-band SBF magnitudes fainter as indicated with an arrow.  
At 5 Gyr, however, the overshooting effects become minimal.  
\label{fig06}}
\end{figure}

\begin{figure}
\epsscale{.8}
\plotone{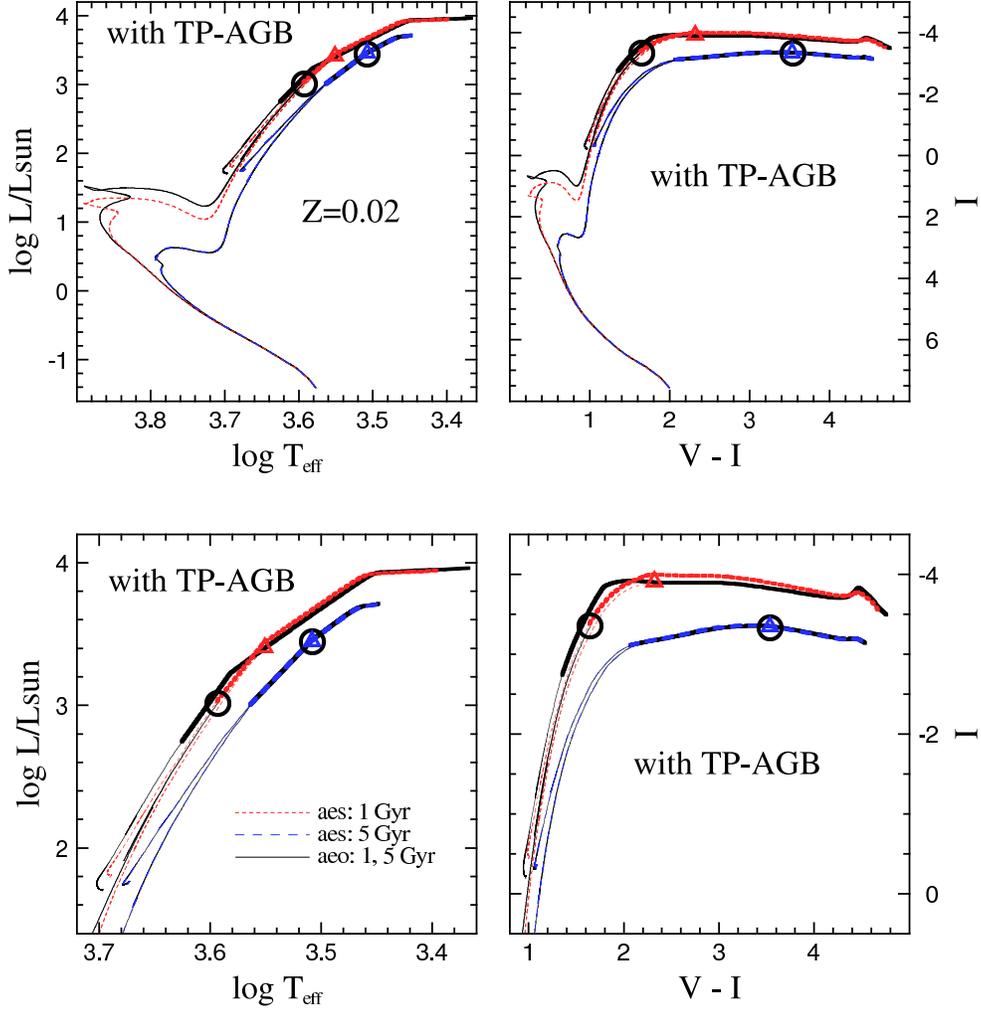}
\caption{Similar to the right panels of Figures 1 and 2, 
but the aes (dashed lines; without convective core overshooting) and 
the aeo (solid lines; with convective core overshooting) Teramo BaSTI 
isochrones are compared in the HR diagrams (left panels) and C-M diagrams 
(right panels) at 1 and 5 Gyr at $Z$=0.02.  RGB tips are denoted with 
triangles for the aes and circles for the aeo, respectively.  Bottom 
panels show the details in the giant branches.  It is 
interesting to note that the aeo models (1) are hotter at the upper MS 
compared to the aes models and (2) do not go all the way to the RGB tip 
at 1 Gyr.  They become virtually identical at 5 Gyr.  The overshooting 
effects on the integrated $V - I$ colors and the $I$-band SBF magnitudes 
that we described in Figure 6 can be understood from this C-M diagram 
(see text).  
\label{fig07}}
\end{figure}

\begin{figure}
\epsscale{1.}
\plotone{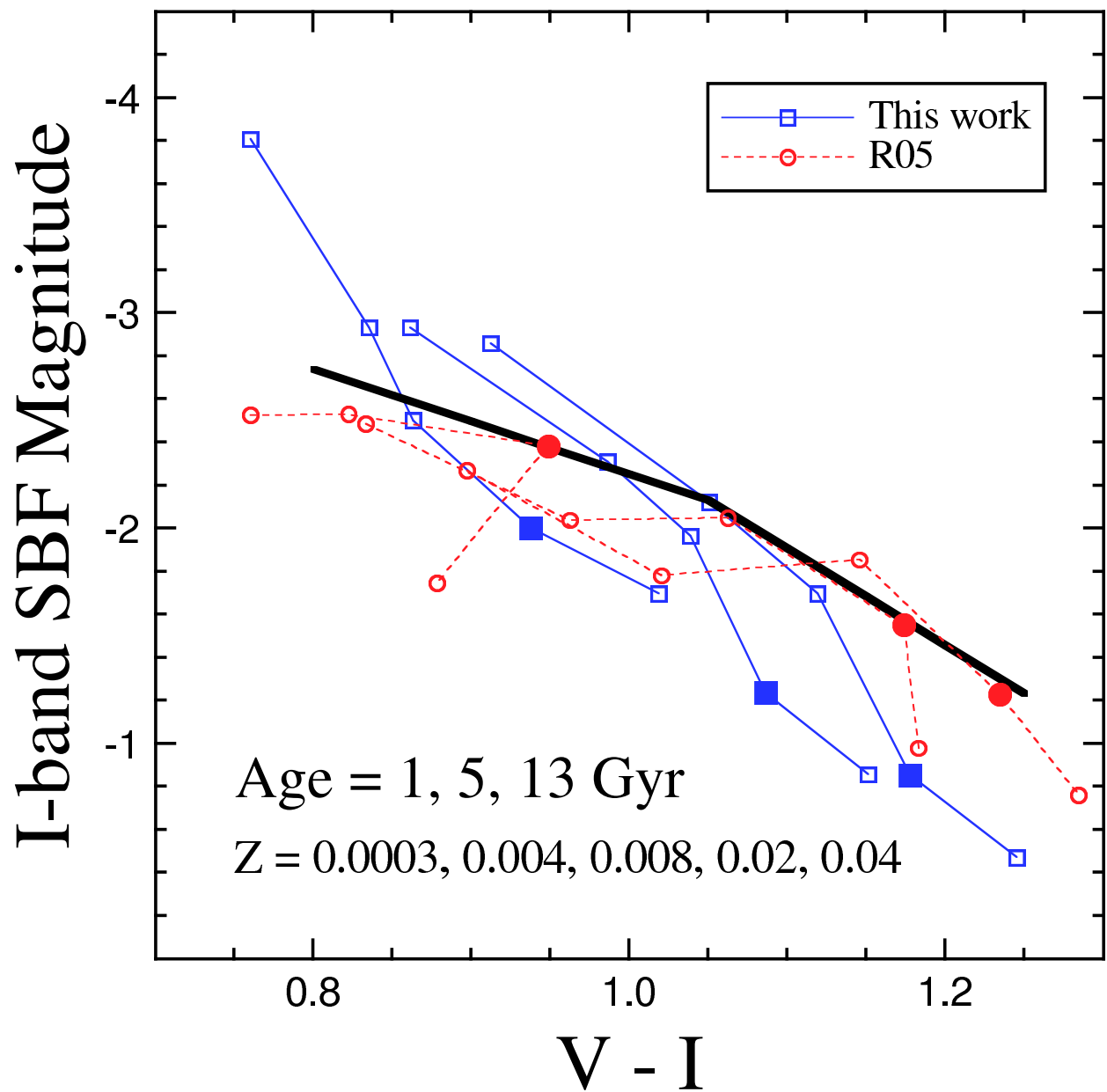}
\caption{Similar to the left panel of Figure 4, but here our solar-scaled 
(sss; without convective core overshooting) $I$-band SBF models 
as a function of $V - I$ are compared with the Teramo SPoT models 
(Raimondo et al. 2005; R05) at 1, 5, and 13 Gyr for five different 
metallicities.  To guide the eye, 
solar-metallicity ($Z$=0.02) is depicted with filled symbols.
\label{fig08}}
\end{figure}

\begin{figure}
\epsscale{1.}
\plotone{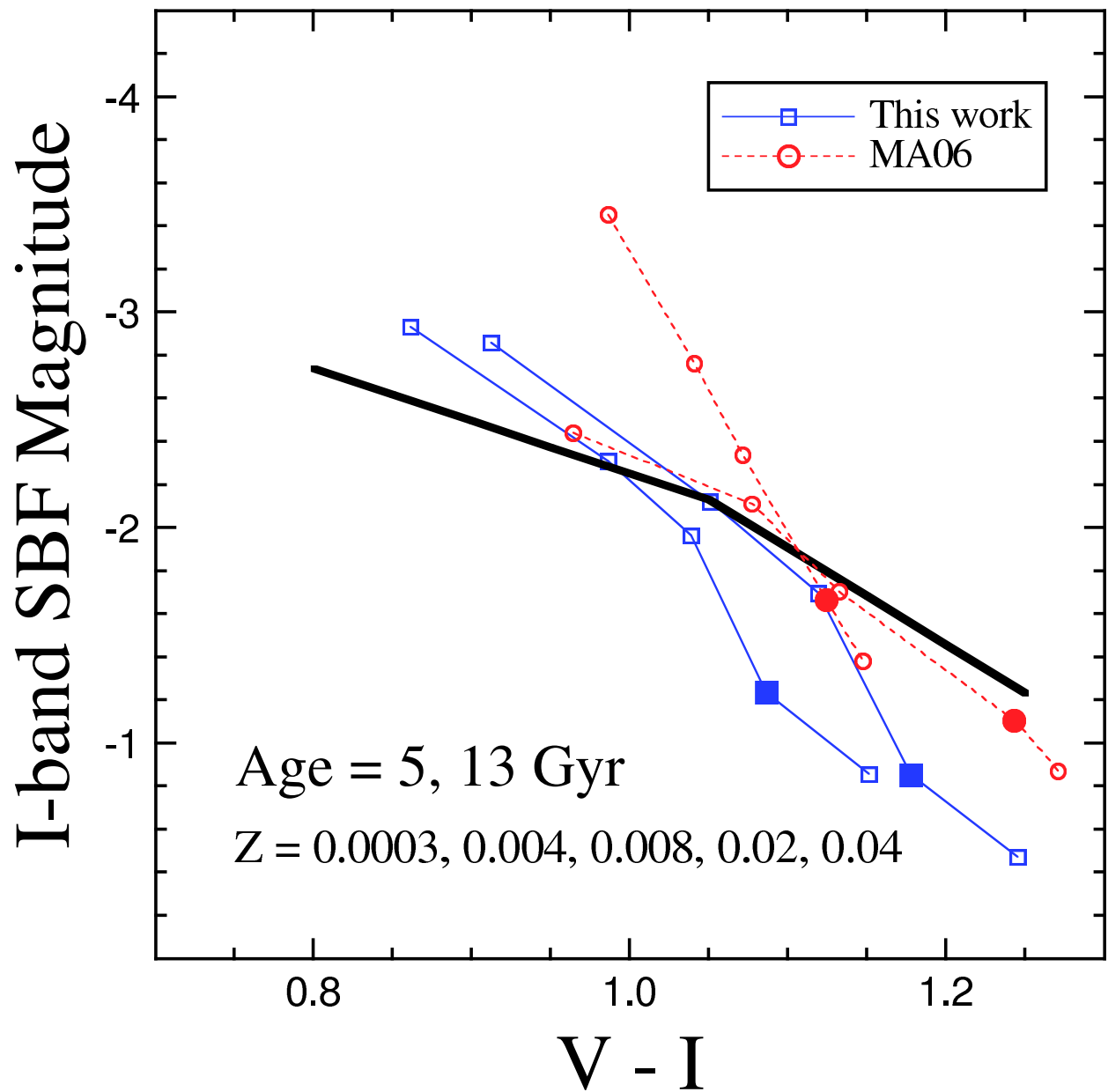}
\caption{Similar to Figure 8, but here our solar-scaled $I$-band SBF models 
as a function of $V - I$ are compared from that of Mar\'in-Franch \& Aparicio 
(2006; MA06) at 5, and 13 Gyr for five different metallicities.  
To guide the eye, solar-metallicity ($Z$=0.02) 
is depicted with filled symbols.
\label{fig09}}
\end{figure}

\begin{figure}
\epsscale{1.}
\plotone{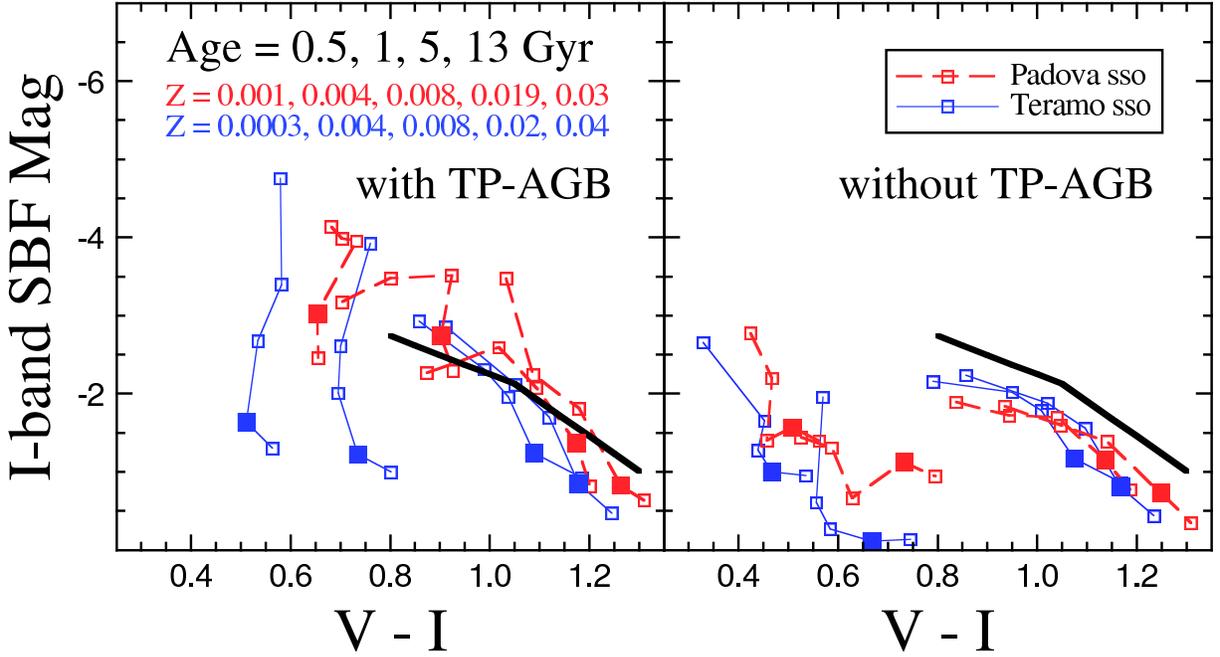}
\caption{Similar to Figure 4, but here we contrast the $I$-band SBF models 
as a function of integrated $V - I$ colors at the same {\em solar-scaled} 
compositions (sso; convective core overshooting is ``on") using the Padova 
(dashed lines) and the Teramo BaSTI (solid lines) stellar models at given 
ages and metallicities.  To guide the eye, solar-metallicity is depicted with 
filled squares.  The observational fiducial lines are same as in Figure 4.  
In general, $V - I$ colors employing the Padova isochrones are 
comparatively redder than that employing the Teramo BaSTI.  
It is noted from the left panel that the $I$-band SBF 
magnitudes based upon the Teramo BaSTI are more than 1 mag fainter 
at younger ages (t $<$ 5 Gyr) with $Z$ $>$ 0.004 
compared to that employing Padova stellar models.  The model disparities 
based on the two different 
stellar models linger even without TP-AGBs as shown on the right panel.  
It is evident though that the inclusion of TP-AGB stages is crucial to match 
the observations, which are the bent thick lines.
\label{fig10}}
\end{figure}

\begin{figure}
\epsscale{.8}
\plotone{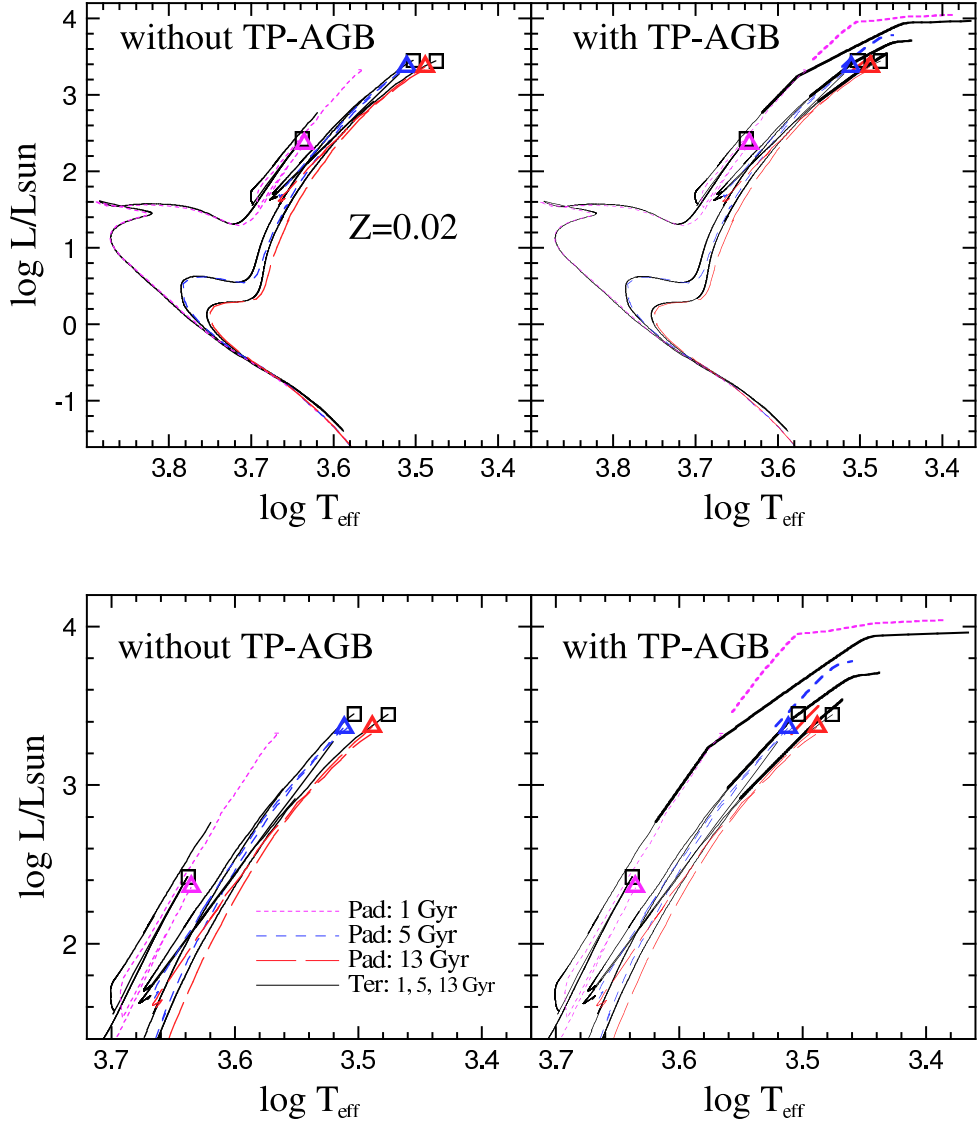}
\caption{Comparison of the Teramo BaSTI (solid lines) and the Padova 
(dashed lines) isochrones at the same {\em solar-scaled} composition 
(sso; convective core overshooting is ``on") in the HR diagrams 
without (left panels) and with (right panels) TP-AGB stages.  
At solar metallicity, three ages (1, 5, 13 Gyr) 
are compared.  RGB tips are marked with squares for Teramo BaSTI and 
triangles for Padova, respectively.  Bottom panels show 
the giant branches in details.  Note that the RGB temperatures 
from the Padova stellar models are generally cooler 
than that from the Teramo BaSTI models and they become 
redder $V - I$ colors as shown in Figure 12.  
\label{fig11}}
\end{figure}

\begin{figure}
\epsscale{.8}
\plotone{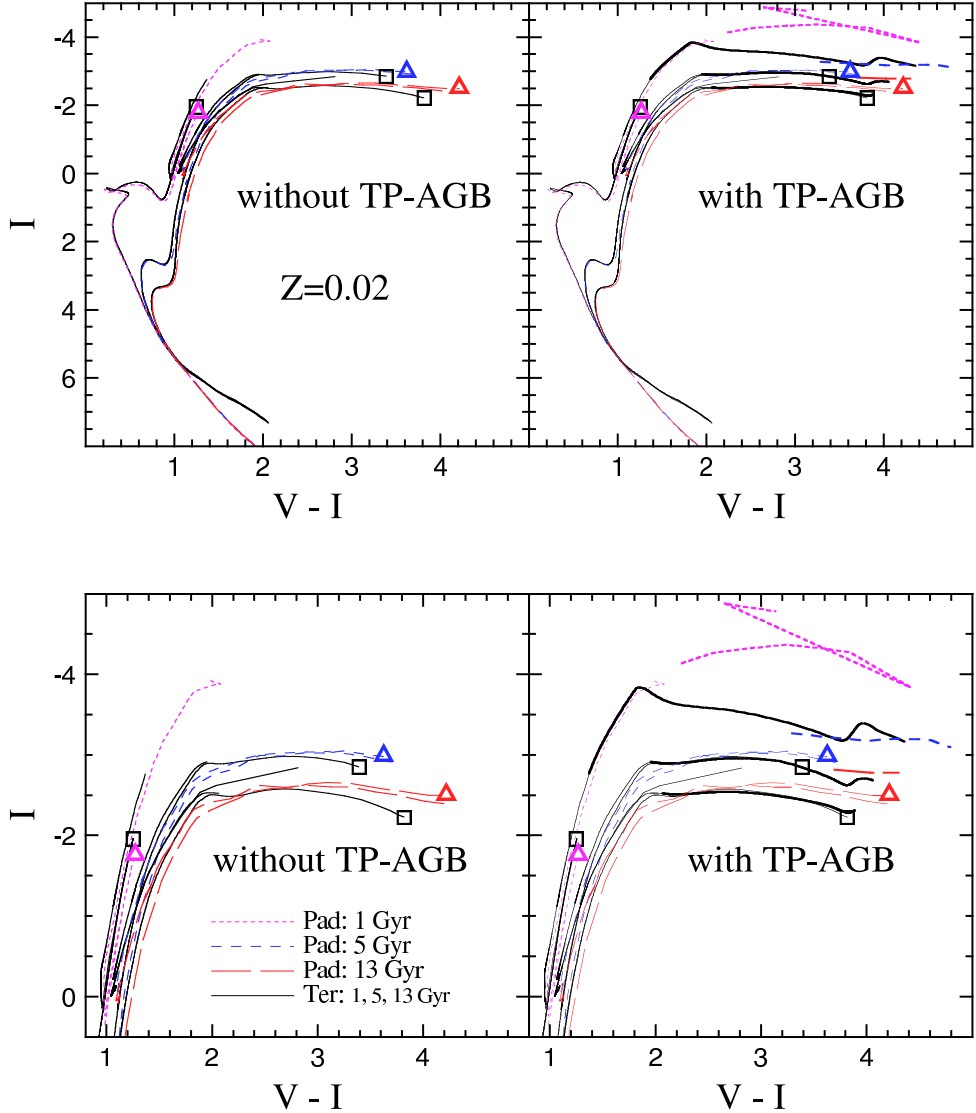}
\caption{Similar to Figure 11, but here displaying the comparison of the 
Teramo BaSTI (solid lines) and the Padova (dashed lines) isochrones in 
$V - I$ vs.\ $I$ C-M diagrams.  Symbols for the RGB tips are same as 
in Figure 11.  Bottom panels show the giant branches in details.  
Note that the Padova giant branches, especially at the upper part, 
are relatively redder than the Teramo BaSTI ones in this 
$V - I$ vs.\ $I$ C-M diagrams.  Note also from the right panels that 
there are some discontinuities for the Padova stellar models at the onset 
of TP-AGBs because of their structural changes (see text).  
\label{fig12}}
\end{figure}

\begin{figure}
\epsscale{1.}
\plotone{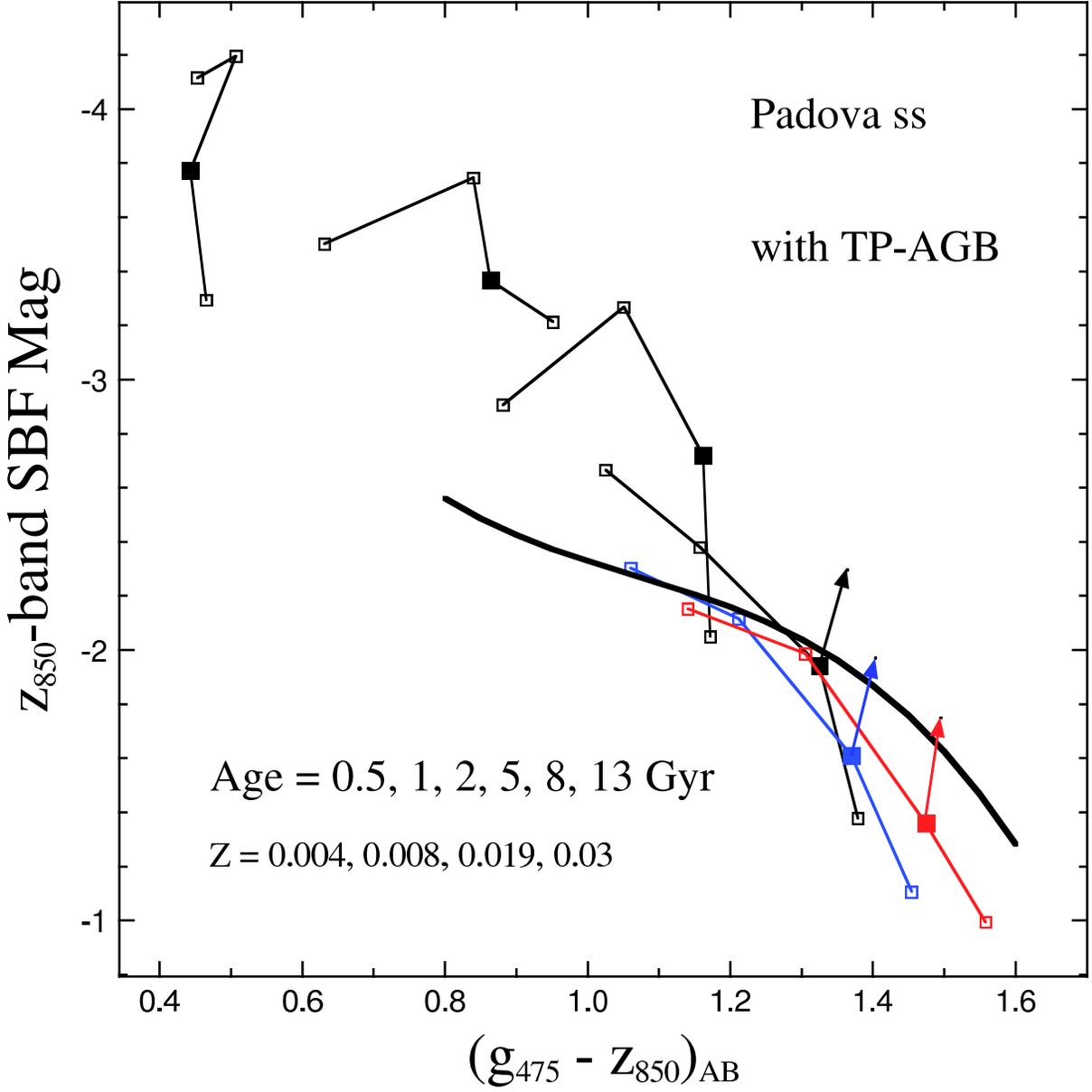}
\caption{Similar to the left panel of Figure 10, but here we contrast 
the {\em HST} ACS/WFC $z_{850}$-band SBF models employing the Padova stellar 
models as a function of integrated $g_{475} - z_{850}$ colors at given ages 
and metallicities.  The magnitudes are all in AB system.  To guide the eye, 
solar metallicities are marked with filled squares and 8 and 13 Gyr are 
displayed in different colors.  The thick curved line is the empirical 
relation from Blakeslee et al.\ (2009).  The arrows at 5, 8, and 13 Gyr 
indicate the estimated 0.4 dex $\alpha$-enhancement effects (see text).
\label{fig13}}
\end{figure}

\begin{figure}
\epsscale{1.}
\plotone{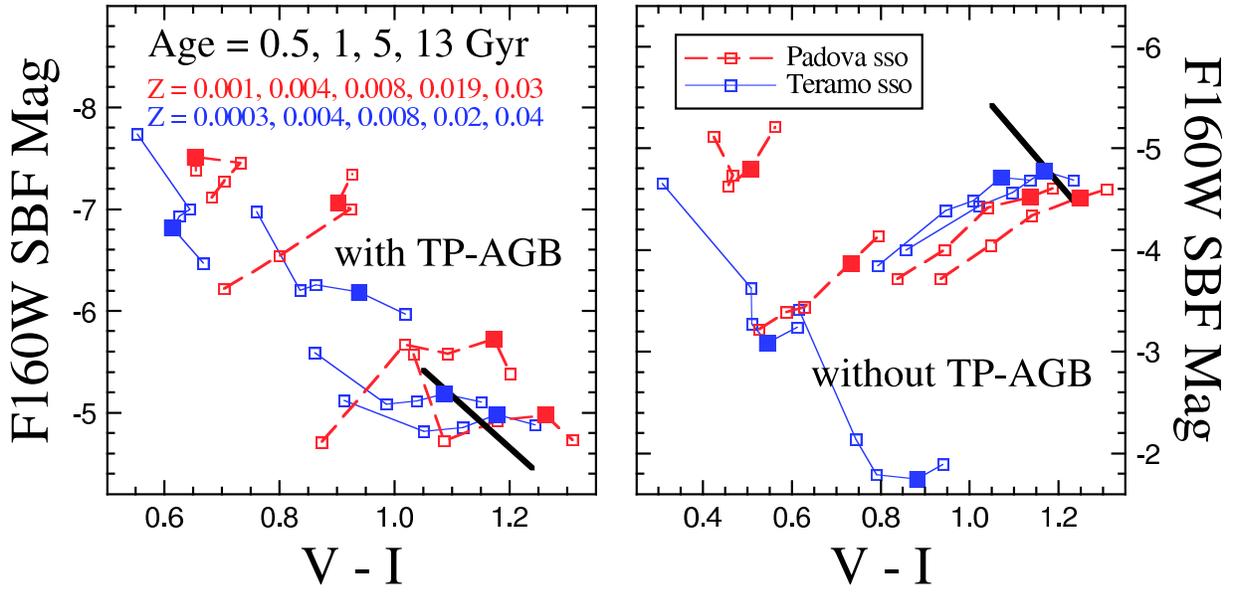}
\caption{Similar to Figure 10, but here the {\em HST} NICMOS 
F160W-band SBF models as a function of integrated $V - I$ colors are 
contrasted using the Padova (dashed lines) and the Teramo BaSTI (solid 
lines) stellar models at the same {\em solar-scaled} compositions 
(sso; convective core overshooting is ``on").  The thicker line is the 
empirical sequence from Jensen et al. (2003).  The importance of the 
inclusion of TP-AGB stages is considerably greater here for the near-IR 
SBF models in order to match the observations.  
It is noted from the left panel that the predicted 
near-IR SBF magnitudes become brighter with bluer colors, as observed, 
when the TP-AGB phase is included.  In the right panel, the opposite 
dependence is seen from the models without the TP-AGB.  
\label{fig14}}
\end{figure}

\begin{figure}
\epsscale{1.}
\plotone{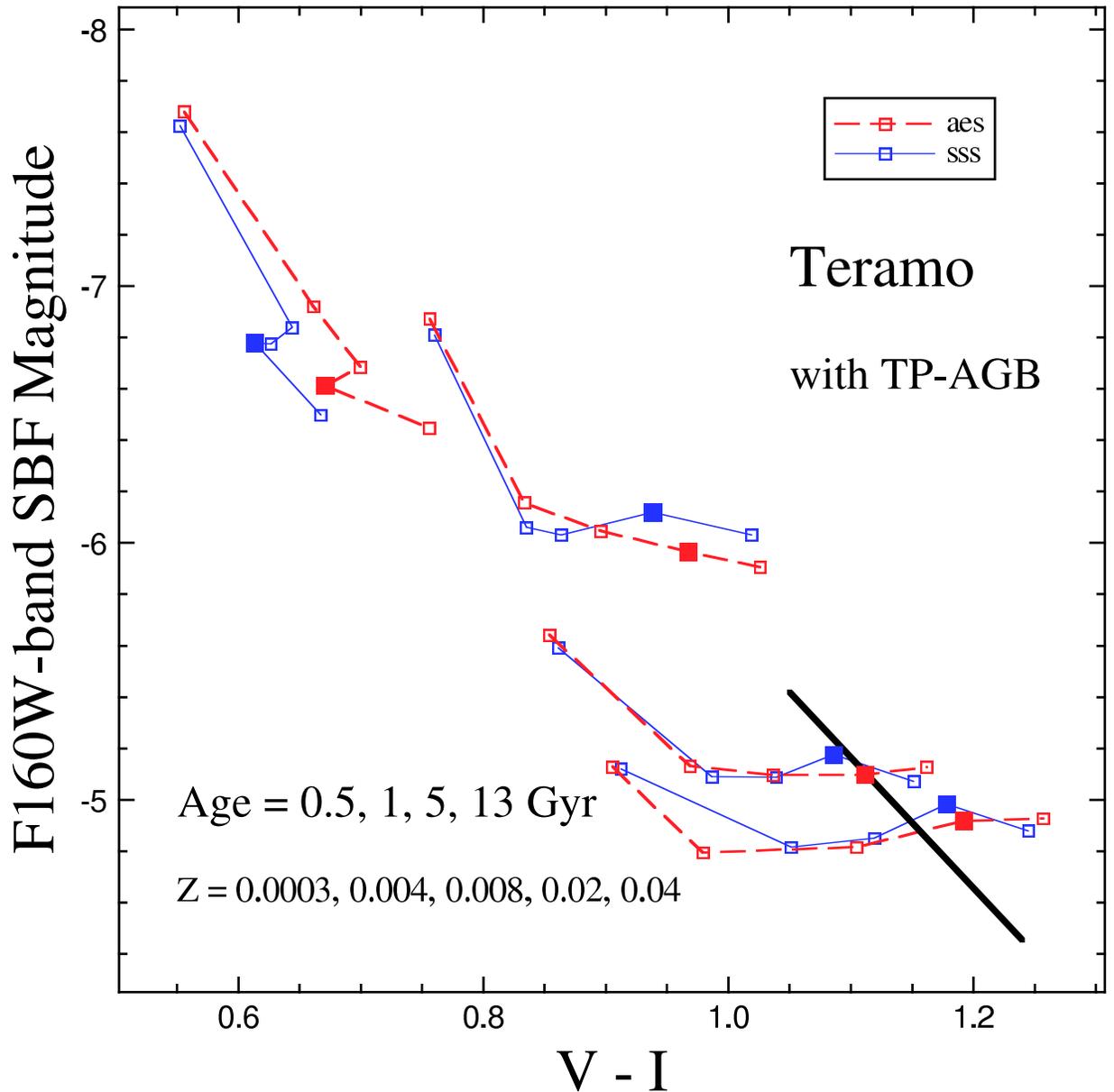}
\caption{Similar to the left panel of Figure 14, but here 
the $\alpha$-element enhancement effects are shown for the {\em HST} 
NICMOS F160W-band SBF magnitudes as a function of integrated $V - I$ 
colors employing the Teramo BaSTI stellar models.  
Solid lines with squares are solar-scaled models, 
while dashed lines with squares are $\alpha$-enhanced models.  
The thick observational fiducial line is same as in Figure 14.  
The F160W-band SBF magnitude differences between the solar-scaled 
and the $\alpha$-enhanced models employing the Teramo BaSTI stellar models 
are relatively small, less than 0.2 mag.  
\label{fig15}}
\end{figure}

\end{document}